\documentclass[12pt]{iopart}

 \usepackage{iopams}
  
 \usepackage{graphicx}
 \usepackage{cite}
 \usepackage{mathrsfs}
  
\begin{document}

\title{A diagrammatic treatment of neutrino oscillations}

\author{D V Naumov$^1$ and V A Naumov$^2$}

\address{$^1$ Dzhelepov Laboratory of Nuclear Problems,
               Joint Institute for Nuclear Research, RU-141980 Dubna, Russia}
  \ead{naumov@nusun.jinr.ru}

\address{$^2$ Bogoliubov Laboratory of Theoretical Physics,
               Joint Institute for Nuclear Research, RU-141980 Dubna, Russia}
  \ead{vnaumov@theor.jinr.ru}

\begin{abstract}

We present a covariant wave-packet approach to neutrino flavor transitions in vacuum.
The approach is based on the technique of macroscopic Feynman diagrams describing
the lepton number violating processes of production and absorption of virtual massive neutrinos
at the macroscopically separated spacetime regions (``source'' and ``detector'').
Accordingly, the flavor transitions are a result of interference of the diagrams 
with neutrinos of different masses in the intermediate states.
The statistically averaged probability of the process is representable as a multidimensional integral
of the product of the factors which describe the differential flux density of massless neutrinos from the source,
differential cross section of the neutrino interaction with the detector and a dimensionless factor responsible
for the flavor transition.
The conditions are analyzed under which the last factor can be treated as the flavor transition
probability in the usual quantum mechanical sense.


\end{abstract}


\vspace*{40mm}
\noindent
{\small 
Published in: {\it J.\ Phys.\ G: Nucl.\ Part.\ Phys.} {\bf 37} (2010) 105014 \\
\phantom{Published in:} http://iopscience.iop.org/0954-3899/37/10/105014 \\
\phantom{Published in:} doi: 10.1088/0954-3899/37/10/105014
}

\maketitle

\section{Introduction}

The neutrino-oscillation hypothesis suggested by Bruno Pontecorvo~\cite{Pontecorvo:57} offers
the most probable and promising explanation of the anomalies discovered in many experiments
with solar, atmospheric, reactor and accelerator neutrinos and antineutrinos.
However the conventional quantum-mechanical (QM) theory of neutrino oscillations faces a number
of methodological difficulties and paradoxes which have been widely discussed by many authors
(see, e.g., \cite{Kayser:81,Giunti:91,Giunti:93,Rich:93,Giunti-Kim:07,Akhmedov:09,Akhmedov:10a} 
and numerous references therein).
It has been recognized that, remaining within the framework of quantum mechanics, some problems of the conventional
theory can be resolved by using wave packets to describe the neutrino states and/or the states of the particles
involved in the processes of neutrino production and detection.
In this way it has been found, e.g., that neutrino oscillations vanish if the neutrinos propagate
over the distances much larger than the coherence length (for in-depth discussion, see \cite{Giunti-Kim:07}).

A more radical approach to the problem is based on quantum field theory
(QFT)~\cite{Grimus:96,Grimus:99,Cardall:99b,Ioannisian:98,Beuthe};
see also \cite{Giunti-Kim:07,Akhmedov:09,Akhmedov:10a} for further references.
In this approach, the sequential processes of neutrino production and detection are treated as a single process
which can be described by the $S$-matrix formalism of QFT.
The peculiarity (or better to say novelty) of this approach is in the use of Feynman diagrams with vertices
separated by macroscopically large spacetime intervals.
Besides, as in the QM case, for the description of the asymptotically free ``in'' and ``out'' states
one has to use wave packets rather than the ordinary (in QFT) Fock states with definite 4-momenta.
In this framework, the neutrinos appear as the propagators of the neutrino mass eigenfields, $\nu_i$, connecting the
macroscopically distant parts of the Feynman diagram.
In the new approach, there is no need in even mentioning the flavor neutrino states or fields,
and the neutrino oscillation phenomenon is a reflection of the usual interference of the Feynman diagrams
with different $\nu_i$'s in the intermediate states.
Such treatment not only enables us to reproduce the standard QM formula for the flavor transition
probability and obtain the pertinent QFT corrections (thus offering a potential possibility for an experimental
verification of the theory) but also impels us to re-examine the concept of neutrino flavor mixing.

The authors of the mentioned pioneer papers do use (explicitly or implicitly) the noncovariant, usually Gaussian
wave packets, which are only applicable, as it will be shown in the next section, to nonrelativistic particles.
However, in the reactions of neutrino production and detection taking part in the current neutrino oscillation
experiments, relativistic particles are as a rule involved.%
\footnote{An important exception from the rule might be experiments with M\"{o}ssbauer antineutrinos
          which are under active discussion in the current literature~
          \cite{Raghavan:05-06,Akhmedov:08-09,Bilenky:08-09,Kopp:09,Potzel:09}.}
Moreover, by an appropriate Lorentz boost, any particle can be made to be ultrarelativistic.
Therefore it is necessary to work out a relativistically covariant (inertial frame-independent) theory of wave packets
applicable for the description of particles with arbitrary momenta.

In this paper we propose a simple covariant wave-packet theory (section~\ref{sec:Wavepackets_theory}),
which is then applied to the calculation of the interference of macroscopic diagrams with massive neutrino exchange
(section~\ref{sec:amplitude}).
We perform a statistical averaging of the probability (the squared absolute value of the overall amplitude) which
leads to a rather general formula for the neutrino event rate in the detector (section~\ref{sec:probability});
the conditions of applicability of the conventional QM formula for the flavor transition probability
are also discussed. In section~\ref{sec:conclusions} we summarise the main results of the formalism.

\section{\label{sec:Wavepackets_theory} Relativistic wave packets}

\subsection{\label{sec:General_consideration} General consideration}

The $S$-matrix formalism of QFT usually deals with one-particle Fock states
\begin{eqnarray}
\label{One-particleState}
|\boldsymbol{k},s\rangle=\sqrt{2E_{\boldsymbol{k}}}\,{a^\dagger_{\boldsymbol{k}s}}|0\rangle
\end{eqnarray}
as asymptotically free states of the fields.
Here $|0\rangle$ is the Fock vacuum state ($a_{\boldsymbol{k}s}|0\rangle=0$), $a_{\boldsymbol{k}s}^\dagger$ and $a_{\boldsymbol{k}s}$
are the creation and annihilation operators of a particle with 3-momentum $\boldsymbol{k}$ and spin projection $s$,
$E_{\boldsymbol{k}}=\sqrt{\boldsymbol{k}^2+m^2}$ and $m$ is the mass of the particle.
The conventional (anti)commutation relations hold:
\begin{eqnarray*}
\{a_{\boldsymbol{q}r},a_{\boldsymbol{k}s}\}=\{a_{\boldsymbol{q}r}^\dagger,a_{\boldsymbol{k}s}^\dagger\}=0,
\qquad
\{a_{\boldsymbol{q}r},a_{\boldsymbol{k}s}^\dagger\}=(2\pi)^3\delta_{sr}\delta\left(\boldsymbol{k}-\boldsymbol{q}\right),
\end{eqnarray*}
where the curly brackets denote a commutator (for boson fields) or anticommutator (for fermion fields).
The Lorentz-invariant normalization of the states \eref{One-particleState} is therefore singular since
\begin{eqnarray*}
\label{PlaneWaveNormalization}
\langle \boldsymbol{q},r|\boldsymbol{k},s\rangle = (2\pi)^3 2E_{\boldsymbol{k}} \delta_{sr}\delta\left(\boldsymbol{k}-\boldsymbol{q}\right).
\end{eqnarray*}
Let us define the wave-packet state localized in a spacetime point $x=(x_0,\boldsymbol{x})$ as a linear superposition of
the one-particle states \eref{One-particleState}
\begin{eqnarray}
\label{WavePacketState}
|\boldsymbol{p},s,x\rangle = \int\frac{d\boldsymbol{k}\,\phi(\boldsymbol{k},\boldsymbol{p})e^{i(k-p)x}}
{(2\pi)^32E_{\boldsymbol{k}}}|\boldsymbol{k},s\rangle,
\end{eqnarray}
in which $\phi(\boldsymbol{k},\boldsymbol{p})$ is a Lorentz-invariant function.
The last requirement implies that the proper Lorentz transformation $\Lambda$
of the 4-vectors $p=(p_0,\boldsymbol{p})$ and  $x=(x_0,\boldsymbol{x})$,
\begin{eqnarray*}
p \longmapsto p'={\Lambda}p,
\qquad 
x \longmapsto x'={\Lambda}x,
\end{eqnarray*}
induces the following unitary transformation $U_{\Lambda}$ of the wave-packet state:
\begin{eqnarray*}
\label{U_transformation_of_k-state}
|\boldsymbol{p},s,x\rangle \longmapsto U_{\Lambda}|\boldsymbol{p},s,x\rangle=|\boldsymbol{p}',s,x'\rangle,
\end{eqnarray*}
assuming that the axis of spin quantization is oriented along the boost or rotation axis.
We see that this transformation rule is in fact the same as that for the Fock states (see, e.g., \cite{Peskin:95})
and this is just the reason of the form of definition given by \eref{WavePacketState}.

We will suppose that the function $\phi(\boldsymbol{k},\boldsymbol{p})$ has a unique sharp maximum at $\boldsymbol{k}=\boldsymbol{p}$,
and its behaviour around the maximum is governed by a small parameter $\sigma$ of the dimension of mass.%
\footnote{In general, the function $\phi(\boldsymbol{k},\boldsymbol{p})$ may involve a finite or infinite set of parameters.
          Here, solely for simplicity, we assume that
          $\left[d\ln\phi(\boldsymbol{k},\boldsymbol{p})/d(k-p)^2\right]_{\boldsymbol{k}=\boldsymbol{p}}\propto\sigma^{-2}>0$.
          Then $\sigma$ is the only essential combination of the mentioned parameters.}
We further require that the wave-packet state \eref{WavePacketState} passes into the Fock state \eref{One-particleState}
as $\sigma\to0$. This is possible if 
\begin{eqnarray}
\label{Correspondence_Principle}
\lim_{\sigma\to0}\phi(\boldsymbol{k},\boldsymbol{p})=(2\pi)^32E_{\boldsymbol{p}}\delta(\boldsymbol{k}-\boldsymbol{p}).
\end{eqnarray}
This \emph{correspondence principle} enables us to impose on the function $\phi(\boldsymbol{k},\boldsymbol{p})$ the following
Lorenz-invariant condition:
\begin{eqnarray}
\label{Normalization_of_phi}
\int\frac{d\boldsymbol{k}\,\phi(\boldsymbol{k},\boldsymbol{p})}{(2\pi)^32E_{\boldsymbol{k}}}=
\int\frac{d\boldsymbol{k}\,\phi(\boldsymbol{k},\boldsymbol{0})}{(2\pi)^32E_{\boldsymbol{k}}}=1.
\end{eqnarray}
Indeed, the dimensionless Lorenz-invariant integral~\eref{Normalization_of_phi} does not depend on $\boldsymbol{p}$.
Hence it can only depend on the dimensionless ratio $\sigma/m$. Owing to~\eref{Correspondence_Principle} the
integral~\eref{Normalization_of_phi} tends to 1 as $\sigma\to0$; thus it is natural to put it equal to one also at finite
(small) $\sigma$. 
Finally it can be said that the form factor function $\phi(\boldsymbol{k},\boldsymbol{p})$ represents, up to a multiplier,
a ``smeared'' $\delta$ function in the momentum space.

Now, we find the wavefunction describing the state~\eref{WavePacketState} in the configuration space.
Let us do this for a spin-1/2 fermion with the field operator 
\begin{eqnarray*}
\Psi(x) = \int \frac{d\boldsymbol{k}}{(2\pi)^3\sqrt{2E_{\boldsymbol{k}}}}
\sum_s\left[a_{\boldsymbol{k}s}u_s(\boldsymbol{k})e^{-ikx}+b_{\boldsymbol{k}s}^{\dagger}v_s(\boldsymbol{k})e^{ikx}\right].
\end{eqnarray*}
The state $\langle 0|\Psi(x)$ can be treated as a linear superposition of one-particle states with definite momenta
at the point $x$ which cannot be characterized by a specific momentum.  
The coordinate representation of the Fock state~\eref{One-particleState} is given by its projection onto the state $\langle 0|\Psi(x)$,
\begin{eqnarray}
\label{PlaneWave}
\langle{0}|\Psi(x)|\boldsymbol{p},s\rangle = u_s(\boldsymbol{p})e^{-ipx},
\end{eqnarray}
which is a plane wave uniformly distributed over the whole spacetime.
In contrast, the wave-packet state~\eref{WavePacketState} is characterized by a momentum distribution governed by
the function $\phi(\boldsymbol{k},\boldsymbol{p})$ concentrated near the most probable momentum $\boldsymbol{p}$.
Besides that, the spacetime distribution of the wave packet is not uniform as can be seen from
the analog of equation~\eref{PlaneWave} for the state~\eref{WavePacketState}:
\begin{eqnarray}
\fl\langle{0}|\Psi(x)|\boldsymbol{p},s,y\rangle
& = e^{-ipy}\left[u_s(\boldsymbol{p})-\nabla_{\boldsymbol{p}}u_s(\boldsymbol{p})
    \cdot\left(i\nabla_{\boldsymbol{x}}+\boldsymbol{p}\right)+\ldots\right]
    \psi(\boldsymbol{p},y-x), \cr
\label{psi_meaning_fermion_c}
& \approx e^{-ipy}u_s(\boldsymbol{p})\psi(\boldsymbol{p},y-x).
\end{eqnarray}
Here we have introduced the Lorentz-invariant function
\begin{eqnarray}
\label{psi(p,x)}
\fl\psi(\boldsymbol{p},x)
= \int\frac{d\boldsymbol{k}}{(2\pi)^32E_{\boldsymbol{k}}}\phi(\boldsymbol{k},\boldsymbol{p})e^{ikx}
= \int\frac{d\boldsymbol{k}}{(2\pi)^32E_{\boldsymbol{k}}}\phi(\boldsymbol{k},\boldsymbol{0})e^{ikx_\star}
= \psi(\boldsymbol{0},x_\star),
\end{eqnarray}
which satisfies the Klein-Gordon equation and describes the wave packet in the coordinate representation.
The last equality in \eref{psi(p,x)} is written in the rest frame of the packet ($\boldsymbol{p}_\star=0$);
the 4-vector $x_\star=(x^0_\star,\boldsymbol{x}_\star)$ is related to $x=(x^0,\boldsymbol{x})$ through the Lorentz boost
along the direction $-\boldsymbol{p}$. 
Since $\psi(\boldsymbol{0},x_\star)$ is an even function of $\boldsymbol{x}_\star$,
it can depend only on the variables $x^0_\star$ and $|\boldsymbol{x}_\star|$, related to the invariant quantities 
$(px)=mx^0_\star$ and $x^2= x_\star^2$.
The approximation~\eref{psi_meaning_fermion_c} is valid under the condition
\begin{eqnarray*}
\left|i\nabla_{\boldsymbol{x}}\ln\psi(\boldsymbol{p},y-x)+\boldsymbol{p}\right|\ll2E_{\boldsymbol{p}},
\end{eqnarray*}
which is fully consistent with other approximations in the subsequent analysis.
Note that the approximate equality~\eref{psi_meaning_fermion_c} becomes exact for (pseudo)scalar fields.
In fact, with a certain loss of simplicity and universality of the formalism, it would be possible to completely
get rid of this approximation if dealing with a spinor wavefunction instead of the scalar function \eref{psi(p,x)}.

From~\eref{psi(p,x)} and \eref{Correspondence_Principle} it follows that $\psi(\boldsymbol{p},x)\to e^{ipx}$ as $\sigma\to 0$,
i.e., as expected, a wave packet utterly localized in the momentum space is delocalized (spread-out) over the configuration
space. The inner product of the states~\eref{WavePacketState} is nonsingular and is given by 
\begin{eqnarray}
\label{ScalarProduct}
\langle\boldsymbol{q},r,y|\boldsymbol{p},s,x\rangle
=\delta_{sr}e^{i\left(qy-px\right)}\mathcal{D}(\boldsymbol{p},\boldsymbol{q};x-y),
\end{eqnarray}
where
\begin{eqnarray}
\label{Dfunction}
\mathcal{D}(\boldsymbol{p},\boldsymbol{q};x)=
\int\frac{d\boldsymbol{k}}{(2\pi)^32E_{\boldsymbol{k}}}\,\phi(\boldsymbol{k},\boldsymbol{p})\phi^*(\boldsymbol{k},\boldsymbol{q})e^{ikx}.
\end{eqnarray}
From~\eref{ScalarProduct} and \eref{Dfunction} it follows that the normalization of the state \eref{WavePacketState} is also finite:
\begin{eqnarray}
\label{V_star_definition_2}
\langle\boldsymbol{p},s,x|\boldsymbol{p},s,x\rangle
  = \mathcal{D}(\boldsymbol{p},\boldsymbol{p};0)
  = 2\overline{E}_{\boldsymbol{p}}\mathrm{V}(\boldsymbol{p}).
\end{eqnarray}
The quantities $\overline{E}_{\boldsymbol{p}}$ and $\mathrm{V}(\boldsymbol{p})$ in~\eref{V_star_definition_2} are, respectively,
the mean energy and effective spatial volume of the packet, defined by
\begin{eqnarray}
\label{MeanEnergy}
\overline{E}_{\boldsymbol{p}} 
= \frac{\int d\boldsymbol{x}\psi(\boldsymbol{p},x)i\partial_0\psi^*(\boldsymbol{p},x)}
  {\int d\boldsymbol{x}|\psi(\boldsymbol{p},x)|^2}
= \frac{1}{\mathrm{V}(\boldsymbol{p})}\int\frac{d\boldsymbol{k}|\phi(\boldsymbol{k},\boldsymbol{p})|^2}{4(2\pi)^3E_{\boldsymbol{k}}},
\end{eqnarray}
\begin{eqnarray}
\label{EffectiveVolume}
\mathrm{V}(\boldsymbol{p})
= \int d\boldsymbol{x}|\psi(\boldsymbol{p},x)|^2
= \int\frac{d\boldsymbol{k}}{(2\pi)^3}\frac{|\phi(\boldsymbol{k},\boldsymbol{p})|^2}{(2E_{\boldsymbol{k}})^2}
= \frac{\mathrm{V}(\boldsymbol{0})}{\Gamma_{\boldsymbol{p}}},
\end{eqnarray}
where $\Gamma_{\boldsymbol{p}} = E_{\boldsymbol{p}}/m$.
So, both $\overline{E}_{\boldsymbol{p}}$ and $\mathrm{V}(\boldsymbol{p})$, as well as the mean momentum $\overline{\boldsymbol{p}}$
defined by a relation similar to \eref{MeanEnergy}, are integrals of motion. It is easy to prove that the mean position of the packet
follows the classical trajectory:
\begin{eqnarray}
\label{MeanPosition_final}
\overline{\boldsymbol{x}} 
= \frac{1}{\mathrm{V}(\boldsymbol{p})} \int d\boldsymbol{x}\psi^*(\boldsymbol{p},x)\boldsymbol{x}\psi(\boldsymbol{p},x)
= \boldsymbol{v}_{\boldsymbol{p}}x_0.
\end{eqnarray}
Here $\boldsymbol{v}_{\boldsymbol{p}}=\overline{\boldsymbol{p}}/\overline{E}_{\boldsymbol{p}}$ is the mean group velocity of the packet,
which coincides with the most probable velocity $\nabla_{\boldsymbol{p}}E_{\boldsymbol{p}}=\boldsymbol{p}/E_{\boldsymbol{p}}$.

\begin{figure}[htb]
\centering
\includegraphics[width=0.8\linewidth]{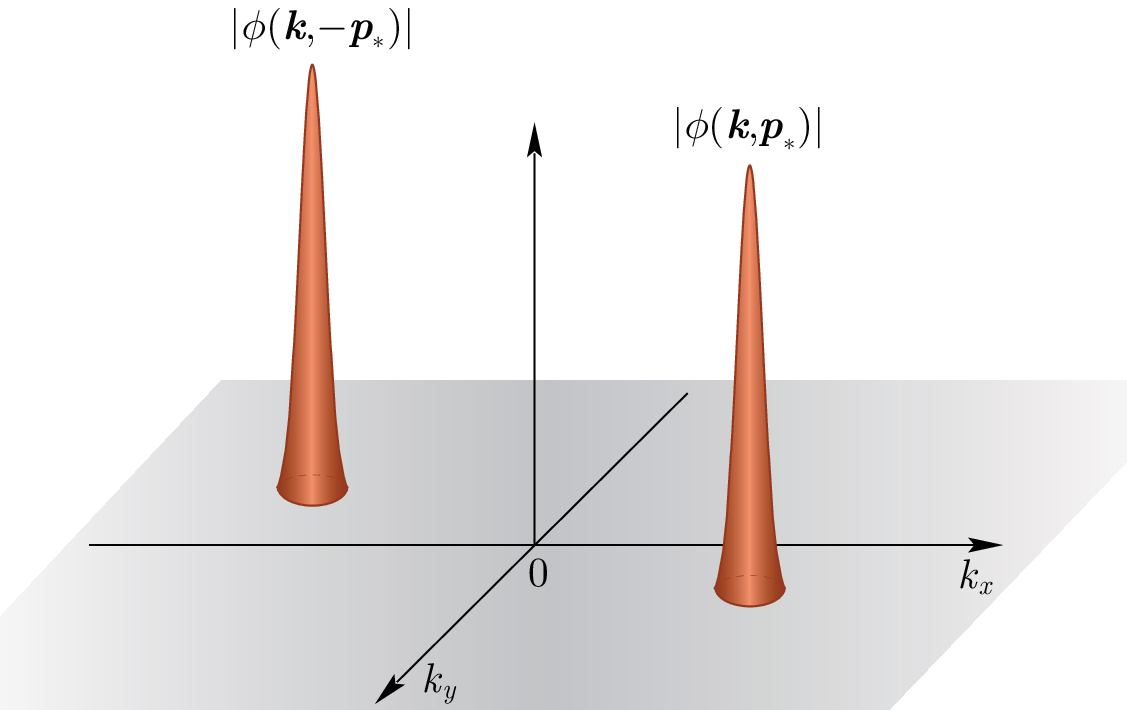}
\protect\caption[Schematic illustration of the vanishing of the integrand in equation~\protect\eref{Dfunction_CIF}
                at large $|\boldsymbol{p}_*|$.]
                {Schematic illustration of the vanishing of the integrand in equation~\protect\eref{Dfunction_CIF}
                at large $|\boldsymbol{p}_*|$.}
\label{fig:Commutator}
\end{figure}

Certain properties of the function~\eref{Dfunction} become especially transparent in the center-of-inertia frame
of the two packets (defined by the condition $\boldsymbol{p}_*+\boldsymbol{q}_*=0$ and denoted from here on by an 
asterisk subscript). In this frame
\begin{eqnarray}
\label{Dfunction_CIF}
\mathcal{D}(\boldsymbol{p}_*,-\boldsymbol{p}_*;x_*-y_*)=
\int\frac{d\boldsymbol{k}}{(2\pi)^32E_{\boldsymbol{k}}}\,\phi(\boldsymbol{k},\boldsymbol{p}_*)
\phi^*(\boldsymbol{k},-\boldsymbol{p}_*)e^{ik(x_*-y_*)}.
\end{eqnarray}
Due to the assumed sharp maximum of $\phi(\boldsymbol{k},\boldsymbol{p})$ at $\boldsymbol{k}=\boldsymbol{p}$, one may expect that
the function \eref{Dfunction_CIF} has a sharp maximum at $\boldsymbol{p}_*=0$ (that is at $\boldsymbol{q}=\boldsymbol{p}$)
and quickly vanishes at large values of $|\boldsymbol{p}_*|$ since the maxima of the multiplicands $|\phi(\boldsymbol{k},\boldsymbol{p}_*)|$
and $|\phi(\boldsymbol{k},-\boldsymbol{p}_*)|$ in the integrand are widely separated in this case and thus their product is small
for \emph{any} $\boldsymbol{k}$, as is illustrated in figure~\ref{fig:Commutator}.
One can see in addition that  the integral \eref{Dfunction_CIF} vanishes at any $|\boldsymbol{p}_*|$ if the points $x_*$ and $y_*$
are sufficiently separated in space (namely, if $|\boldsymbol{x}_*-\boldsymbol{y}_*|\gg1/\sigma$) since the phase factor
$e^{-i\boldsymbol{k}(\boldsymbol{x}_*-\boldsymbol{y}_*)}$ in the integrand of \eref{Dfunction_CIF} rapidly oscillates in this case.

\subsection{\label{sec:RGP} Relativistic Gaussian packets}

In further consideration we will use a simple model of the state \eref{WavePacketState}, relativistic Gaussian packet (RGP),
for which the function $\phi(\boldsymbol{k},\boldsymbol{p})$ is of the form
\begin{eqnarray}
\label{phi_RG_final(a)}
\phi(\boldsymbol{k},\boldsymbol{p}) = \frac{2\pi^2}{\sigma^2K_1(m^2/2\sigma^2)}
\exp\left(-\frac{E_{\boldsymbol{k}}E_{\boldsymbol{p}}-\boldsymbol{k}\boldsymbol{p}}{2\sigma^2}\right)
~\stackrel{\mathrm{def}}{=}~\phi_G(\boldsymbol{k},\boldsymbol{p}),
\end{eqnarray}
where $K_1(t)$ is the modified Bessel function of the third kind of order 1. One can verify that the function \eref{phi_RG_final(a)}
has the correct plane-wave limit~\eref{Correspondence_Principle} and satisfies the condition~\eref{Normalization_of_phi}.
In what follows we assume $\sigma^2 \ll m^2$. Then the function \eref{phi_RG_final(a)} can be rewritten as an asymptotic 
expansion by a small parameter $\sigma^2/m^2$:
\begin{eqnarray}
\label{phi_RG_final(b)}
\phi_G(\boldsymbol{k},\boldsymbol{p}) = \frac{2\pi^{3/2}}{\sigma^2}\frac{m}{\sigma}
\exp\left[\frac{(k-p)^2}{4\sigma^2}\right]      
\left[1+\frac{3\sigma^2}{4m^2}+\mathcal{O}\left(\frac{\sigma^4}{m^4}\right)\right].
\end{eqnarray}
It is easy to convince that the nonrelativistic limit of the function \eref{phi_RG_final(b)} coincides, up to a normalization
factor, with the usual (noncovariant) Gaussian distribution
$\varphi_G(\boldsymbol{k}-\boldsymbol{p}) \propto \exp{\left[-(\boldsymbol{k}-\boldsymbol{p})^2/4\sigma^2\right]}$
widely used in the literature. However at relativistic momenta the functions $\phi_G$ and $\varphi_G$ significantly differ
from each other. For example, in the vicinity of the maximum $\boldsymbol{k}=\boldsymbol{p}$,
\begin{eqnarray*}
\label{phi_RG_final(c)}
\phi_G(\boldsymbol{k},\boldsymbol{p}) \approx \frac{2\pi^{3/2}}{\sigma^2}\frac{m}{\sigma}
\exp\left[-\frac{\left(\boldsymbol{k}-\boldsymbol{p}\right)^2}{4\sigma^2\Gamma_{\boldsymbol{p}}^2}\right]
\qquad (\boldsymbol{k}\sim\boldsymbol{p}).
\end{eqnarray*}
We see that in this case the relativistic effect consists in a ``renormalization'' of the wave-packet width
($\sigma\to\sigma\Gamma_{\boldsymbol{p}}$). This renormalization is very essential for the neutrino production
and detection processes involving relativistic particles.

\begin{figure}[htb]
\centering
\includegraphics[width=0.6\linewidth,clip=true]{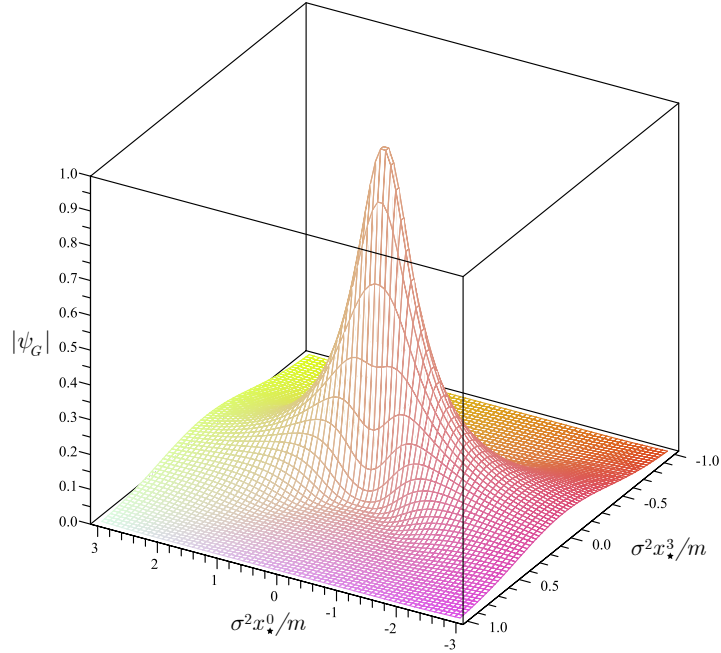}
\protect\caption[A 3D plot of $|\psi_G(\boldsymbol{0},x_\star)|$ as a function of $\sigma^2x_\star^0/m$ and $\sigma^2|\boldsymbol{x}_\star/m$.]
                {A 3D plot of $|\psi_G(\boldsymbol{0},x_\star)|$ as a function of $\sigma^2x_\star^0/m$ and $\sigma^2x_\star^3/m$
                (assuming that $\boldsymbol{x}_\star=(0,0,x_\star^3)$). The calculations are done for $\sigma/m=0.1$.}
\label{fig:mod_psi}
\end{figure}

The coordinate representation of the RGP wavefunction is found to be
\begin{eqnarray}
\label{psi_RGP_exact}
\psi_G(\boldsymbol{p},x)=\frac{K_1({\zeta}m^2/2\sigma^2)}{{\zeta}K_1(m^2/2\sigma^2)},
\end{eqnarray}
where $\zeta$ is the complex dimensionless scalar variable:
\begin{eqnarray*}
\zeta = \sqrt{1-\frac{4\sigma^2}{m^2}\left[\sigma^2x^2+i(px)\right]}.
\end{eqnarray*}
Here and below the square root means the principal square root.
Figure~\ref{fig:mod_psi} shows the shape of the function $|\psi_G(\boldsymbol{0},x_\star)|$ calculated, as an example, for $\sigma/m=0.1$
and plotted as a function of the dimensionless variables $\sigma^2x_\star^0/m$ and $\sigma^2x_\star^3/m$, assuming that $\boldsymbol{x}_\star$
is directed along the third axis; of course, $|\psi_G(\boldsymbol{0},x_\star)|$ is an even function of both variables.
As one can observe, $|\psi_G(\boldsymbol{0},x_\star)|$ rapidly vanishes with increase of $x_\star^3$ and its spatial width grows
with $x_\star^0$, i.e. RGP spreads with time as any wave packet with nonzero mass. Nonetheless the effective volume \eref{EffectiveVolume}
does not depend on time since the spreading of $|\psi_G(\boldsymbol{0},x_\star)|$ precisely compensates its fall-off and the normalization
of the RGP state remains constant.

In the RGP model, the function \eref{Dfunction} which defines the scalar product of the states \eref{ScalarProduct} reads
\begin{eqnarray}
\label{D_RGP_exact}
\mathcal{D}(\boldsymbol{p},\boldsymbol{q};x)
  =    2\overline{E}_{\boldsymbol{p}}\mathrm{V}(\boldsymbol{p})\frac{K_1(zm^2/\sigma^2)}{zK_1(m^2/\sigma^2)}
~\stackrel{\mathrm{def}}{=}~\mathcal{D}_G(\boldsymbol{p},\boldsymbol{q};x);
\end{eqnarray}
the complex dimensionless scalar variable $z$ in \eref{D_RGP_exact} is defined by
\begin{eqnarray*}
z = \frac{1}{2m}\sqrt{(p+q)^2-4\sigma^2\left[\sigma^2x^2+i(p+q)x\right]}.
\end{eqnarray*}

Now we have to determine the physical conditions under which the spreading of RGP can be neglected, since just in this regime
the wave packets can naturally be associated with the (quasi)stable particles and then used (instead of the plane waves)
as the asymptotically free states of incoming and outgoing fields in the $S$-matrix formalism.

\subsection{\label{sec:Nondiffluent_regime} Nondiffluent regime}

An accurate analysis of the asymptotic expansion of $\ln\left[\psi_G(\boldsymbol{0},x_\star)\right]$ in powers of the
small parameter $\sigma^2/(m^2\zeta)$, taking into account the inequalities $|\zeta|\ge1$ and $|\arg\zeta|<\pi/2$,
provides the following (necessary and sufficient) conditions of the nondiffluent behavior:
\begin{eqnarray}
\label{TheRestrictions_psi}
\sigma^2(x^0_\star)^2   \ll m^2/\sigma^2,
\qquad
\sigma^2|\boldsymbol{x}_\star|^2 \ll m^2/\sigma^2.
\end{eqnarray}
They can be rewritten in the equivalent but explicitly Lorentz-invariant form:
\begin{eqnarray}
\label{TheRestrictions_psi_inv}
(px)^2        \ll m^4/\sigma^4,
\qquad
(px)^2-m^2x^2 \ll m^4/\sigma^4.
\end{eqnarray}
Under these conditions, the function \eref{psi_RGP_exact} reduces to the simple and transparent form
\begin{eqnarray}
\label{psi_AsymptoticExpansion_lowest_star}
\fl\psi_G(\boldsymbol{p},x)
= \exp\left(imx^0_\star-\sigma^2\boldsymbol{x}_\star^2\right)
= \exp\left\{i(px)-\frac{\sigma^2}{m^2}\left[(px)^2-m^2x^2\right]\right\}.
\end{eqnarray}
Let us mention some properties of this approximation which will be referred to as the
contracted relativistic Gaussian packet (CRGP).
It is easy to verify that the mean coordinate of the packet follows the classical trajectory~\eref{MeanPosition_final}
and the absolute value of the function $\psi_G(\boldsymbol{p},x)$ is invariant under the transformations $x_0\longmapsto x_0+\tau$,
$\boldsymbol{x}\longmapsto\boldsymbol{x}+\boldsymbol{v}_{\boldsymbol{p}}\tau$. It is also obvious that $|\psi_G(\boldsymbol{p},x)|=1$
along the classical trajectory $\boldsymbol{x}=\boldsymbol{v}_{\boldsymbol{p}}x_0$ and $|\psi_G(\boldsymbol{p},x)|<1$
with any deviation from it.
In the nonrelativistic limit, the wavefunction \eref{psi_AsymptoticExpansion_lowest_star} takes the form
\begin{eqnarray*}
\psi_G(\boldsymbol{p},x)\approx\exp\left[im\left(x_0-\boldsymbol{v}_{\boldsymbol{p}}\boldsymbol{x}\right)
-\sigma^2\left|\boldsymbol{x}-\boldsymbol{v}_{\boldsymbol{p}}x_0\right|^2\right].
\end{eqnarray*}
In the CRGP model one obtains that $\mathrm{V}(\boldsymbol{0})=[\pi/(2\sigma^2)]^{3/2}\equiv\mathrm{V}_{\star}$ and
$\overline{E}_{\boldsymbol{p}}=E_{\boldsymbol{p}}$.

The CRGP approximation for the function~\eref{D_RGP_exact} can be derived by analyzing the asymptotic expansion
of its logarithm in powers of the parameter $\sigma^2/(m^2z)$ and taking into account the inequalities $|z|\ge1$ and
$|\arg\,z|<\pi/2$. This yields
\begin{eqnarray}
\label{D_AsymptoticExpansion_b}
\fl\mathcal{D}_G(\boldsymbol{p}_*,-\boldsymbol{p}_*;x_*) = \frac{2m\mathrm{V}_{\star}}{\Gamma_*^{3/2}}
      \exp\left[imx^0_*-\frac{m^2\left(\Gamma_*-1\right)}{\sigma^2}
     -\frac{\sigma^2\boldsymbol{x}_*^2}{2\Gamma_*}\right],
\end{eqnarray}
where $\Gamma_*=E_*/m$ ($E_* \equiv E_{\boldsymbol{p}^*}$). The applicability conditions of this approximation,
$\sigma^2(x^0_*)^2 \ll E_*^2/\sigma^2$ and $\sigma^2|\boldsymbol{x}_*|^2 \ll E_*^2/\sigma^2$,
are found to be fully compatible with the conditions \eref{TheRestrictions_psi}.
As was expected (see section~\ref{sec:RGP}), the function \eref{D_AsymptoticExpansion_b} rapidly vanishes
if either $|\boldsymbol{p}_*|$ or $|\boldsymbol{x}_*|$ (or both) are sufficiently large.
The function $\mathcal{D}_G$ has a number of unobvious {\it a priori} properties.
For example, $|\mathcal{D}_G|$ exponentially vanishes at subrelativistic energies ($\Gamma_*-1 \sim 1$).
At nonrelativistic energies, the function $\mathcal{D}_G$ in the lab.\ frame is given by
\begin{eqnarray*}
\fl\mathcal{D}_G(\boldsymbol{p},\boldsymbol{q};x)
\approx  2m\mathrm{V}_{\star}\exp\left[im\left(x_0-\boldsymbol{v}\boldsymbol{x}\right)
        -\frac{m^2\left|\boldsymbol{v}_{\boldsymbol{p}}-\boldsymbol{v}_{\boldsymbol{q}}\right|^2}{8\sigma^2}
        -\frac{\sigma^2\left|\boldsymbol{x}-\boldsymbol{v}x_0\right|^2}{2}\right],
\end{eqnarray*}
where $\boldsymbol{v}_{\boldsymbol{p}}=\boldsymbol{p}/m$, $\boldsymbol{v}_{\boldsymbol{q}}=\boldsymbol{q}/m$
(assuming that $|\boldsymbol{v}_{\boldsymbol{p},\boldsymbol{q}}|\ll1$)
and $\boldsymbol{v}=\frac{1}{2}\left(\boldsymbol{v}_{\boldsymbol{p}}+\boldsymbol{v}_{\boldsymbol{q}}\right)$.

As an application of equation~\eref{D_AsymptoticExpansion_b}, we consider the norm of a state with two identical
noninteracting packets. 
The following exact model-independent relation holds:
\begin{eqnarray}
\label{M_2_example}
\fl\frac{\langle\boldsymbol{p}_1,s_1,x_1;\boldsymbol{p}_2,s_2,x_2|\boldsymbol{p}_1,s_1,x_1;\boldsymbol{p}_2,s_2,x_2\rangle}
{[2\overline{E}_{\boldsymbol{p}}\mathrm{V}(\boldsymbol{p})]^2}
= 1\pm\delta_{s_1s_2}\frac{|\mathcal{D}(\boldsymbol{p}_1,\boldsymbol{p}_2;x_1-x_2)|^2}
{[2\overline{E}_{\boldsymbol{p}}\mathrm{V}(\boldsymbol{p})]^2},
\end{eqnarray}
where the signs ``$+$'' and ``$-$'' refer to bosons and fermions, respectively.
From \eref{D_AsymptoticExpansion_b} it follows that at $\boldsymbol{p}_1=\boldsymbol{p}_2=0$, the right-hand side
of \eref{M_2_example} is equal to
$1\pm\delta_{s_1s_2}\exp\left(-\sigma^2|\boldsymbol{x}_1-\boldsymbol{x}_2|^2\right)$.
Hence the effects of Bose-Einstein attraction and Pauli repulsion (appearing for $s_1=s_2$)
are essential only if the distance between the centers of the packets is comparable or less than their effective
dimensions ($|\boldsymbol{x}_1-\boldsymbol{x}_2|^2\lesssim\sigma^{-2}$),
that is, exactly  when it is already necessary to take into account the dynamic interactions between the packets.
At sufficiently large distances between the packets, the quantum statistics is of no importance any more.
Similar conclusions can be proved to be valid for the states with arbitrary number of free identical packets.

\section{\label{sec:Overlap_integrals} Overlap integrals}

\begin{figure}[htb]
\centering
\includegraphics[width=0.35\linewidth]{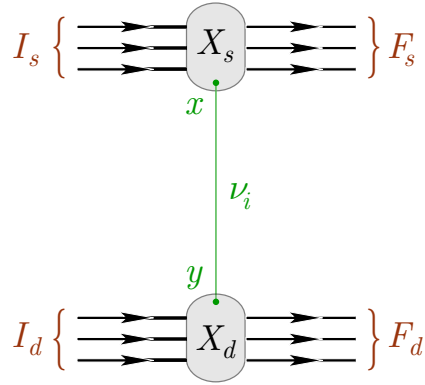}
\protect\caption[A generic macroscopic Feynman diagram with exchange of massive neutrino.]
                {A generic macroscopic Feynman diagram with exchange of massive neutrino.}
\label{fig:Macrograph_Generic}
\end{figure}

We will deal with the Feynman diagrams, the generic structure of which is shown in figure~\ref{fig:Macrograph_Generic}.
The external legs of such diagrams correspond to asymptotically free incoming (``in'') and outgoing (``out'') wave packets
in the coordinate representation, that is, to the wavefunctions $\psi_a\left(\boldsymbol{p}_a,x_a\right)$ and
$\psi_b^*\left(\boldsymbol{p}_b,x_b\right)$ specified by the most probable momenta $\boldsymbol{p}_{a,b}$,
spacetime coordinates $x_{a,b}$, masses $m_{a,b}$ and parameters $\sigma_{a,b}$.
Here and below we use the following notation: $I_s$ ($F_s$) is the set of in (out) packets in the block $X_s$ (``source''),
$I_d$ ($F_d$) is the set of in (out) packets in the block $X_d$ (``detector''). The internal line connecting the blocks
$X_s$ and $X_d$ denotes the causal Green's function of the neutrino mass eigenfield $\nu_i$ ($i=1,2,3$).
The blocks $X_s$ and $X_d$ are assumed to be macroscopically separated in spacetime; this explains the term ``macroscopic diagram''.

In the calculations with the macrodiagrams of such kind, we will encounter the four-dimensional overlap integrals
$\mathbb{V}_s(q)$ and $\mathbb{V}_d(q)$ defined as follows:
\begin{eqnarray*}
\label{OverlapVolumes_def}
\fl\mathbb{V}_{s,d}(q) = \int dx e^{\pm iqx}
\Big[\prod_{a\in{I_{s,d}}}e^{-ip_ax_a}\psi_a  \left(\boldsymbol{p}_a,x_a-x\right)\Big]
\Big[\prod_{b\in{F_{s,d}}}e^{ ip_bx_b}\psi_b^*\left(\boldsymbol{p}_b,x_b-x\right)\Big].
\end{eqnarray*}
In the CRGP approximation, these integrals can be written in the form
\begin{eqnarray}
\label{OverlapVolumes}
\fl\mathbb{V}_{s,d}(q) =  \int dx\exp\Big[i\left(\pm qx-q_{s,d}x\right)
-\sum_{\varkappa{\in}S,D}T_{\varkappa}^{\mu\nu}\left(x_{\varkappa}-x\right)_{\mu}\left(x_{\varkappa}-x\right)_{\nu}\Big],
\end{eqnarray}
where 
\begin{eqnarray*}
q_s=\sum_{a\in{I_s}}p_a-\sum_{b\in{F_s}}p_b,
\qquad
q_d=\sum_{a\in{I_d}}p_a-\sum_{b\in{F_d}}p_b,
\end{eqnarray*}
\begin{eqnarray*}
T_{\varkappa}^{\mu\nu} = \sigma_{\varkappa}^2\left(u_{\varkappa}^{\mu}u_{\varkappa}^{\nu}-g^{\mu\nu}\right),
\qquad
S=I_s{\oplus}F_s,
\qquad
D=I_d{\oplus}F_d,
\end{eqnarray*}
$u_{\varkappa}=p_{\varkappa}/m_{\varkappa}=\Gamma_{\varkappa}(1,\boldsymbol{v}_{\varkappa})$ is the 4-velocity of packet
$\varkappa$ ($\varkappa=a,b$) and, finally, $q$ is the 4-momentum of the virtual neutrino.
Let us now define the symmetric overlap tensors for the source and detector:
\begin{eqnarray*}
\label{Re_sd}
\Re_s^{\mu\nu} = \sum_{\varkappa{\in}S}T_{\varkappa}^{\mu\nu}
\qquad\mathrm{and}\qquad
\Re_d^{\mu\nu} = \sum_{\varkappa{\in}D}T_{\varkappa}^{\mu\nu}.
\end{eqnarray*}
One can prove that these tensors are positive definite provided $\sigma_\varkappa\ne0$ ($\forall\varkappa$);
thus, there exist the positive-definite tensors $\widetilde{\Re}_s^{\mu\nu}$ and $\widetilde{\Re}_d^{\mu\nu}$ such that
$\widetilde{\Re}_{s,d}^{\mu\lambda}\left(\Re_{s,d}\right)_{\lambda\nu}=\delta^{\mu}_{\nu}$ or, in the matrix form,
$\widetilde{\Re}_{s,d}=||\widetilde{\Re}_{s,d}^{\mu\nu}||=g\,\Re_{s,d}^{-1}\,g$, where
$\Re_{s,d}=||\Re_{s,d}^{\mu\nu}||$ and $g=||g_{\mu\nu}||$. Obviously, $|\Re_{s,d}|>0$ and $|\widetilde{\Re}_{s,d}|=|\Re_{s,d}|^{-1}$.
These facts allow us to compute the integrals \eref{OverlapVolumes} explicitly.
For this aim, we define the quantities
\begin{eqnarray}
\label{Delta_sd}
\widetilde{\delta}_{s,d}(K)
 = (4\pi)^{-2}|\Re_{s,d}|^{-1/2}\exp\left(-\frac{1}{4}\widetilde{\Re}_{s,d}^{\mu\nu}K_{\mu}K_{\nu}\right), \\
\label{X_sd}
X_{s,d}^{\mu}
 = \widetilde{\Re}_{s,d}^{\mu\nu}\,\sum_{\varkappa}T_{\varkappa\nu}^{\lambda}x_{\varkappa\lambda}
 = \widetilde{\Re}_{s,d}^{\mu\nu}\,\sum_{\varkappa}\sigma_{\varkappa}^2
   \left[(u_{\varkappa}x_{\varkappa})u_{\varkappa\nu}-x_{\varkappa\nu}\right], \\
\label{S_sd}
\mathfrak{S}_{s,d}
 = \sum_{\varkappa,\varkappa'} \left(\delta_{\varkappa\varkappa'}T_{\varkappa}^{\mu\nu}-T_{\varkappa\mu'}^{\mu}\,
   \widetilde{\Re}_{s,d}^{\mu'\nu'}\,
   T_{\varkappa'\nu'}^{\nu}\right)x_{\varkappa\mu}x_{\varkappa'\nu}
\end{eqnarray}
(here $\varkappa,\varkappa'{\in}S,D$ and $K$ is an arbitrary 4-momentum). With this notation, we find the
following compact expression for the integral \eref{OverlapVolumes}:
\begin{eqnarray*}
\label{V_sd}
\mathbb{V}_{s,d}(q) = (2\pi)^4\widetilde{\delta}_{s,d}\left(q{\mp}q_{s,d}\right)
\exp\left[-\mathfrak{S}_{s,d}~{\pm}~i\left(q{\mp}q_{s,d}\right)X_{s,d}\right].
\end{eqnarray*}
Let us clarify the physical meaning of functions \eref{Delta_sd}--\eref{S_sd}.
In the plane-wave limit ($\sigma_\varkappa\to 0$) the factors $\widetilde{\delta}_s\left(q-q_s\right)$ and
$\widetilde{\delta}_d\left(q+q_d\right)$ become usual $\delta$ functions responsible for the exact energy-momentum
conservation in the source and detector vertices, whereas at nonzero $\sigma_\varkappa$, they only lead
to an approximate energy-momentum conservation: the probability of the process $I_s{\oplus}I_d \to F_s{\oplus}F_d$
is strongly suppressed at a small disbalance in the 4-momenta of the interacting packets and the allowed disbalance
is defined by the tensors $\widetilde{\Re}_s^{\mu\nu}$ and $\widetilde{\Re}_d^{\mu\nu}$, i.e., ultimately, by the momentum
spreads of the packets.

The functions $\exp\left(-\mathfrak{S}_s\right)$ and $\exp\left(-\mathfrak{S}_d\right)$ are the geometric suppression factors
conditioned by a partial overlap of the wave packets in the spacetime region of their interaction.
This can be seen after converting \eref{S_sd} to the form%
\footnote{In this derivation we have used the translation invariance of the functions $\mathfrak{S}_{s,d}$.}
\begin{eqnarray}
\label{mathfrak_S_sd_final_a}
\mathfrak{S}_{s,d}
= \sum_{\varkappa}T_{\varkappa}^{\mu\nu}\left(x_{\varkappa}-X_{s,d}\right)_{\mu}\left(x_{\varkappa}-X_{s,d}\right)_{\nu}
\end{eqnarray}
and taking into account that $\mathfrak{S}_{s,d}$ and $X_{s,d}$ are invariants under the group of transformations
$x_{\varkappa}^0 \longmapsto x_{\varkappa}^0+\tau_{\varkappa}$,
$\boldsymbol{x}_{\varkappa} \longmapsto \boldsymbol{x}_{\varkappa}+\boldsymbol{v}_{\varkappa}\tau_{\varkappa}$
with arbitrary real parameters $\tau_{\varkappa}$.
The latter symmetry allows, for each packet $\varkappa{\in}S,D$ having a nonzero velocity $\boldsymbol{v}_{\varkappa}$,
the vector $\boldsymbol{x}_{\varkappa}$ in \eref{mathfrak_S_sd_final_a} to be replaced with the vector
\begin{eqnarray}
\label{ImpactVector}
\boldsymbol{b}_{\varkappa}
=\left(\boldsymbol{x}_{\varkappa}-\boldsymbol{X}_{s,d}\right)-\left[\boldsymbol{n}_{\varkappa}
 \left(\boldsymbol{x}_{\varkappa}-\boldsymbol{X}_{s,d}\right)\right]\boldsymbol{n}_{\varkappa}
\qquad
(\boldsymbol{n}_{\varkappa} = \boldsymbol{v}_{\varkappa}/|\boldsymbol{v}_{\varkappa}|)
\end{eqnarray}
(whose absolute value,
$|\boldsymbol{b}_{\varkappa}|=\left|\boldsymbol{n}_{\varkappa}\times\left(\boldsymbol{x}_{\varkappa}-\boldsymbol{X}_{s,d}\right)\right|$,
is the minimum distance between the classical world line of the center of the packet $\varkappa$ and the point
$\boldsymbol{X}_{s,d}$), and the zero component $x_{\varkappa}^0$ to be replaced with the time $b_{\varkappa}^0$
of maximum approach of the packet center to the point $\boldsymbol{X}_{s,d}$, this time being equal to
\begin{eqnarray}
\label{ImpactTime}
b_{\varkappa}^0
=\left(x_{\varkappa}^0-X^0_{s,d}\right)-|\boldsymbol{v}_{\varkappa}|^{-1}\boldsymbol{n}_{\varkappa}
 \left(\boldsymbol{x}_{\varkappa}-\boldsymbol{X}_{s,d}\right).
\end{eqnarray}
The 4-vector $b_{\varkappa}=(b_{\varkappa}^0,\boldsymbol{b}_{\varkappa})$, built from \eref{ImpactVector} and \eref{ImpactTime},
is a relativistic analog of the usual impact parameter, so it is natural to call it the impact vector.
The 4-vectors $X_s$ and $X_d$ can be called, accordingly, the impact points.
If $\boldsymbol{v}_{\varkappa}\ne0$ ($\forall\varkappa$) then, after the substitution $x_{\varkappa}~\longmapsto~b_{\varkappa}$,
the expression \eref{mathfrak_S_sd_final_a} becomes
\begin{eqnarray*}
\label{mathfrak_S_sd_final_c}
\mathfrak{S}_{s,d}
= \sum_{\varkappa}\sigma_{\varkappa}^2
  \left[\left(\Gamma_{\varkappa}^2-1\right)\left(b^0_{\varkappa}\right)^2+\boldsymbol{b}_{\varkappa}^2\right]
= \sum_{\varkappa}\sigma_{\varkappa}^2|\boldsymbol{b}_{\varkappa}^\star|^2.
\end{eqnarray*}
In the last equality, the contribution from each packet is written in its own rest frame
(in which $|\boldsymbol{b}_{\varkappa}^\star|=|\boldsymbol{x}_{\varkappa}^\star-\boldsymbol{X}_{s,d}^\star|$),
and hence the temporal limitation $\boldsymbol{v}_{\varkappa}\ne0$ can be bypassed. 
The physical meaning of the factors $\exp\left(-\mathfrak{S}_s\right)$ and $\exp\left(-\mathfrak{S}_d\right)$ is now transparent:
the interaction of in and out packets is unsuppressed ($\mathfrak{S}_{s,d}\ll1)$ if all impact parameters $|\boldsymbol{b}_{\varkappa}^\star|$
are small relative to the effective dimensions of the packets $\varkappa$ ($\sim1/\sigma_{\varkappa}$).
In the lab.\ frame, the suppression of ``unlucky'' configurations of the packets' world lines are defined by both the
space and time components of the impact vectors $b_{\varkappa}$. The contribution of the time components $b^0_{\varkappa}$
is less important for nonrelativistic packets (with $\Gamma_{\varkappa}\sim1$), but is very essential for ultrarelativistic packets
(with $\Gamma_{\varkappa}\gg1$).

Our consideration suggests that the impact points $X_s$ and $X_d$ identify the spacetime position of
the regions of effective interaction of the packets in the source and detector, respectively.
The more intensive interaction of the packets, the closer their world lines are placed with respect to the impact points.
The condition that the interaction regions in the source and detector are macroscopically separated
is equivalent to the macroscopic separation of the impact points.
The world line configurations and impact point coordinates have no concern with the dynamics governed by the interaction Lagrangian,
being uniquely defined by the initial (final) coordinates, group velocities and effective dimensions of the asymptotically
free in (out) packets. But, of course, the full probability of the process $I_s{\oplus}I_d \to F_s{\oplus}F_d$
is defined by both the suppression factors $\exp\left(-\mathfrak{S}_s\right)$ and $\exp\left(-\mathfrak{S}_d\right)$
and interaction dynamics.

It is evident now that the spacetime remoteness of the initial 4-coordinates of the in and out wave packets
from the interaction region means simply that 
\begin{eqnarray*}
x_a^0 & \ll X_{s,d}^0, 
\qquad 
\left|\boldsymbol{x}_a-\boldsymbol{X}_{s,d}\right| \gg {\max}_a\left(\sigma_a^{-1}\right),
\qquad 
a {\in} I_{s,d}, \\
x_b^0 & \gg X_{s,d}^0, 
\qquad 
\left|\boldsymbol{x}_b-\boldsymbol{X}_{s,d}\right| \gg {\max}_b\left(\sigma_b^{-1}\right),
\qquad 
b {\in} F_{s,d}.
\end{eqnarray*}
Therewith, for the applicability of the CRGP model it is necessary that the components $X_{s,d}^\mu-x_{\varkappa}^\mu$
remain finite in magnitude and satisfy the conditions \eref{TheRestrictions_psi_inv}.
It is meaningful to note that the requirement of spatial remoteness is in fact unnecessary
for specification of the asymptotically free initial and final states.
Indeed, some of the packets (e.g. a target nucleus or decaying meson in the source vertex)
can be at rest in the lab.\ frame and thus they \emph{must} be spatially close to the corresponding
impact point; otherwise, the resultant interaction amplitude will be strongly suppressed by the geometric factor
$\exp\left(-\mathfrak{S}_s\right)$ or $\exp\left(-\mathfrak{S}_d\right)$.

\section{\label{sec:amplitude} Calculation of a macroscopic amplitude}

\begin{figure}[htb]
\centering
\includegraphics[width=0.4\linewidth]{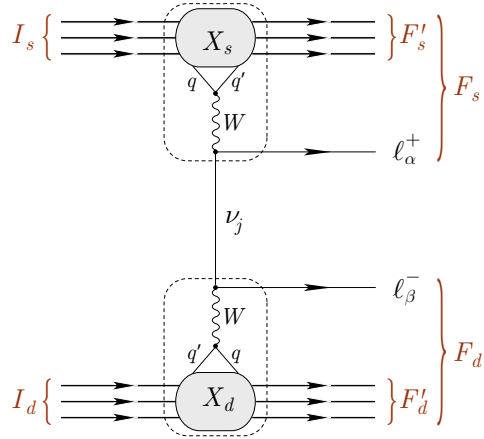}
\protect\caption[A macrodiagram describing the process \protect$I_s {\oplus} I_d \to F_s'+\ell_\alpha~{\oplus}~F_d'+\ell_\beta$.]
                {A macroscopic Feynman diagram describing the process \protect\eref{Macroprocess_A}.}
\label{fig:Macrograph_Class_A}
\end{figure}

As a practically important example, we consider the charged-current induced production of the
charged leptons $\ell_\alpha^+$ and $\ell_\beta^-$ ($\ell_{\alpha,\beta}=e,\mu,\tau$) in the process
\begin{eqnarray}
\label{Macroprocess_A}
I_s {\oplus} I_d \to F_s'+\ell_\alpha^+~{\oplus}~F_d'+\ell_\beta^-.
\end{eqnarray}
We assume for definiteness that all the external substates $I_s$, $I_d$, $F_s'$ and $F_d'$ consist exclusively
of (asymptotically free) hadronic wave packets.
Consequently, if $\alpha\ne\beta$, the process~\eref{Macroprocess_A} violates the lepton numbers $L_\alpha$ and $L_\beta$ 
that is only possible  via exchange of massive neutrinos (no matter whether they are Dirac or Majorana particles).
In the lowest nonvanishing order in electroweak interactions, the process~\eref{Macroprocess_A} is described
by the sum of diagrams of figure~\ref{fig:Macrograph_Class_A}.
Let $X_s$ and $X_d$ be the impact points defined by eq.~\eref{X_sd}.
We require these points and by that the effective regions of interactions in the source and detector (the areas symbolically outlined
in figure~\ref{fig:Macrograph_Class_A} by dashed curves) to be macroscopically separated and the conditions
\begin{eqnarray*}
x_a^0 \ll X_{s,d}^0 \quad (\forall a \in I_{s,d})
\quad\mathrm{and}\quad
x_b^0 \gg X_{s,d}^0 \quad (\forall b \in F_{s,d})
\end{eqnarray*}
to be fulfilled.
Hence the incoming and outgoing states are thought to be direct products of free one-packet states \eref{WavePacketState},
each normalized according to the relation \eref{V_star_definition_2}.

In the framework of the standard model (SM) phenomenologically extended by inclusion of a neutrino mass term, the quark-lepton blocks of the
diagram of figure~\ref{fig:Macrograph_Class_A} are described by the Lagrangian
\begin{eqnarray*}
\mathcal{L}_W(x)=-\frac{g}{2\sqrt{2}}\left[j_\ell(x)W(x)+j_q(x)W(x)+\mathrm{H.c.}\right],
\end{eqnarray*}
where $g$ is the $SU(2)$ gauge coupling constant, $j_\ell$ and $j_q$ are the lepton and quark weak charged currents,
\begin{eqnarray*}
j_\ell^{\mu}(x) = \sum_{{\alpha}i}V_{{\alpha}i}^*\,\overline{\nu}_i(x)O^\mu\ell_{\alpha}(x),
\qquad
j_q^{\mu}(x) = \sum_{qq'}V_{qq'}^{'*}\,\overline{q}(x)O^\mu q'(x),
\end{eqnarray*}
$V_{{\alpha}i}$ ($\alpha=e,\mu,\tau$, $i=1,2,3$) and $V'_{qq'}$ ($q=u,c,t$, $q'=d,s,b$) are the elements of the neutrino and quark
mixing matrices ($\boldsymbol{V}$ and $\boldsymbol{V}'$, respectively), $\ell_{\alpha}(x)$ and $\ell_{\beta}(x)$ are the charged
lepton fields, and $O^\mu=\gamma^\mu(1-\gamma_5)$. The customary notation is used for other fields and Dirac $\gamma$-matrices.
The normalized dimensionless amplitude of the process~\eref{Macroprocess_A}
\begin{eqnarray*}
\langle\mathrm{\bf out}|\mathbb{S}|\mathrm{\bf in}\rangle
\left(\langle\mathrm{\bf in}|\mathrm{\bf in}\rangle\langle\mathrm{\bf out}|\mathrm{\bf out}\rangle\right)^{-1/2}
~\stackrel{\mathrm{def}}{=}~\mathcal{A}_{\beta\alpha}
\end{eqnarray*}
is given by the fourth order of the perturbation theory in the coupling constant $g$:
\begin{eqnarray}
\label{AmplitudeDef}  
\fl \mathcal{A}_{\beta\alpha}
= \frac{1}{\mathcal{N}}\left(\frac{-ig}{2\sqrt{2}}\right)^4\langle F_s{\oplus}F_d|
     T\int dxdx'dydy':j_\ell(x)W(x):\,:j_q(x')W^\dag(x'): \cr
  \times:j_\ell^\dag(y)W^\dag(y):\,:j_q^\dag(y')W(y'):\mathbb{S}_h|I_s{\oplus}I_d \rangle.
\end{eqnarray}
Here $\mathbb{S}_h = \exp\left[i\int dz\mathcal{L}_{\mathrm{h}}(z)\right]$, $\mathcal{L}_{\mathrm{h}}(z)$ is the Lagrangian
of strong and electromagnetic interactions responsible for nonperturbative processes of fragmentation and hadronization;
$T$ and $:\ldots:$ are the standard symbols for the chronological and normal ordering of local operators. The normalization factor
$\mathcal{N}$ in the CRGP approximation is given by
\begin{eqnarray}
\label{NormalizationFactor}
\mathcal{N}^2 = \langle\mathrm{\bf in}|\mathrm{\bf in}\rangle\langle\mathrm{\bf out}|\mathrm{\bf out}\rangle
              = \prod_{\varkappa{\in}I_s{\oplus}I_d{\oplus}F_s{\oplus}F_d}2E_{\varkappa}\mathrm{V}_{\varkappa}(\boldsymbol{p}_{\varkappa}).
\end{eqnarray}
The assumed narrowness of the wave packets in the momentum space, macroscopic remoteness of the interaction regions in
the source and detector, and the consideration of translation invariance allow us to represent the hadronic part of
the amplitude \eref{AmplitudeDef} in the factorized form
\begin{eqnarray*}
\fl\langle F_s'{\oplus}F_d'|T\left[:j_q^{\mu}(x):\,:j_q^{\dag\nu}(y):\mathbb{S}_h\right]|I_s{\oplus}I_d\rangle 
= \mathcal{J}_{s}^{\mu }(p_S)\mathcal{J}_{d}^{\nu*}(p_D)\Pi',
\end{eqnarray*}
\begin{eqnarray*}
\fl\Pi' =       \Big[\prod_{a{\in}I_s }e^{-ip_ax_a}\psi_a  (\boldsymbol{p}_a,x_a-x)\Big]
                \Big[\prod_{b{\in}F_s'}e^{ ip_bx_b}\psi_b^*(\boldsymbol{p}_b,x_b-x)\Big] \cr
          \times\Big[\prod_{a{\in}I_d }e^{-ip_ax_a}\psi_a  (\boldsymbol{p}_a,x_a-y)\Big] 
                \Big[\prod_{b{\in}F_d'}e^{ ip_bx_b}\psi_b^*(\boldsymbol{p}_b,x_b-y)\Big],
\end{eqnarray*}
where $\mathcal{J}_{s}(p_S)$ and $\mathcal{J}_{d}(p_D)$ are the $c$-number hadronic currents in which the strong interactions
are taken into account nonperturbatively, and $p_S$ and $p_D$ denote the sets of the momentum and spin variables
of the hadronic states. 
Now, by applying Wick's theorem and the known properties of the leptonic wave packets, the amplitude \eref{AmplitudeDef}
can be rewritten in the following way:  
\begin{eqnarray}
\label{Amplitude_2}
\fl\mathcal{A}_{\beta\alpha} = \frac{g^4}{64\mathcal{N}}\sum_j V_{{\beta}j}\mathcal{J}_d^{\nu*}\overline{u}(\boldsymbol{p}_\beta)O^{\nu'}
\mathbb{G}^j_{\nu\nu'\mu'\mu}\left(\{\boldsymbol{p}_{\varkappa},x_{\varkappa}\}\right)
O^{\mu'}v(\boldsymbol{p}_\alpha)\mathcal{J}_s^{\mu}V^*_{{\alpha}j},
\end{eqnarray}
where we have introduced the tensor function
\begin{eqnarray}
\label{GeneralizedGreenFunction}
\fl\mathbb{G}^j_{\nu\nu'\mu'\mu}\left(\{\boldsymbol{p}_{\varkappa},x_{\varkappa}\}\right)
= \int\frac{dq}{(2\pi)^4}\mathbb{V}_d(q)\Delta_{\nu\nu'}(q-p_{\beta})\Delta^j(q)\Delta_{\mu'\mu}(q+p_{\alpha})\mathbb{V}_s(q).
\end{eqnarray}
Here $\mathbb{V}_s(q)$ and $\mathbb{V}_d(q)$ are the overlap integrals discussed in detail in section~\ref{sec:Overlap_integrals},
\begin{eqnarray*}
\Delta^j(q)=i\left(\hat{q}-m_j+i0\right)^{-1}
\end{eqnarray*}
and $\Delta_{\mu\nu}$ are the propagators of, respectively, the massive neutrino $\nu_j$
and $W$ boson (the explicit form of $\Delta_{\mu\nu}$ is not used below), $m_j$ and $m_W$ are their masses,
$v(\boldsymbol{p}_\alpha)$ and $\overline{u}(\boldsymbol{p}_\beta)$ are the Dirac bispinors describing the leptons
$\ell_\alpha^+$ and $\ell_\beta^-$, respectively. Here and below, the spin indices and arguments of the functions
$\mathcal{J}_{s,d}$ are omitted for short.

Let us examine the amplitude \eref{Amplitude_2} at large values of $|\boldsymbol{X}_s-\boldsymbol{X}_d|$.
The integral \eref{GeneralizedGreenFunction} can be evaluated by means of the theorem of Grimus-Stockinger (GS)~\cite{Grimus:96}. 
Let $\Phi=\Phi(\boldsymbol{q})$ be a thrice continuously differentiable function such that $\Phi$ itself and its first
and second derivatives decrease not slowly than $|\boldsymbol{q}|^{-2}$ as $|\boldsymbol{q}|\to\infty$.
Then, according to the GS theorem, in the asymptotic limit of $L=|\boldsymbol{L}|\to\infty$,
\begin{eqnarray*}
\fl\int\frac{d\boldsymbol{q}}{(2\pi)^3}\,
\frac{\Phi(\boldsymbol{q})e^{i\boldsymbol{q}\boldsymbol{L}}}{s-\boldsymbol{q}^2+i0}\sim\left\{
\begin{array}{lll}
-\frac{1}{4{\pi}L}\Phi\left(\sqrt{s}\,\boldsymbol{L}/{L}\right)\exp\left(i\sqrt{s}L\right)
+& \mathcal{O}\left(L^{-3/2}\right)
 & \mathrm{at}~s>0, \\
 &  \mathcal{O}\left(L^{-2}\right)
 & \mathrm{at}~s<0.
\end{array}
\right.
\end{eqnarray*}
The integrand in \eref{GeneralizedGreenFunction} satisfies the formulated requirements.
The integral over $q_0$, which remains after applying the GS theorem, can be evaluated by the regular saddle-point method.
Here we describe only the result of this lengthy calculation performed in the ultrarelativistic approximation,
which means that $q_s^0 \approx -q_d^0 \gg m_j$, $j=1,2,3$.
The stationary saddle point $q_0=E_j$ (which can be naturally treated as the effective energy of the
virtual neutrino $\nu_j$) is found as a series in powers of the small parameter $r_j=m_j^2/(2E_{\nu}^2)$:%
\footnote{We limited ourselves to the first order of the perturbation expansion in $r_j$. However, the next-order corrections
          are, in fact, needed to define properly the range of applicability of the obtained result.}
\begin{eqnarray*}
E_j  = E_{\nu}\left[1-\mathfrak{n}r_j+\mathcal{O}\left(r_j^2\right)\right],
\qquad
E_{\nu}=\frac{Yl}{R},
\qquad
\mathfrak{n}=\frac{\boldsymbol{Y}\boldsymbol{l}}{Yl}, \\
l=(1,\boldsymbol{l}),
\qquad
\boldsymbol{l}=\frac{\boldsymbol{L}}{L},
\qquad
\boldsymbol{L}=\boldsymbol{X}_d-\boldsymbol{X}_s, \\
Y^{\mu}  = \widetilde{\Re}_s^{\mu\nu}q_{s\nu}-\widetilde{\Re}_d^{\mu\nu}q_{d\nu},
\qquad
R=\left(\widetilde{\Re}_s^{\mu\nu}+\widetilde{\Re}_d^{\mu\nu}\right)l_{\mu}l_{\nu}.
\end{eqnarray*}
The effective momentum and velocity of the virtual neutrino are then found as $\boldsymbol{p}_j=P_j\boldsymbol{l}$ and
$\boldsymbol{v}_j=\boldsymbol{p}_j/E_j=v_j\boldsymbol{l}$, respectively, where
\begin{eqnarray*}
\label{P_j,v_j}
P_j  = \sqrt{E_j^2-m_j^2} = E_{\nu}\left[1-\left(\mathfrak{n}+1\right)r_j+\mathcal{O}\left(r_j^2\right)\right]
\end{eqnarray*}
and
\begin{eqnarray*}
v_j = 1-r_j+\mathcal{O}\left(r_j^2\right).
\end{eqnarray*}
As is easy to see, $E_j=P_j=E_\nu=q_s^0=-q_d^0$ in the limit of $m_j=0$ and assuming the exact energy-momentum conservation.
But, in the general case, the effective 4-momentum $p_j=(E_j,\boldsymbol{p}_j)$ is determined by the mean momenta and effective
dimensions of the external wave packets involved in the process \eref{Macroprocess_A}. Finally, by introducing the notation
\begin{eqnarray}
\label{Omega_j}
\Omega_j(T,L) = i\left(E_jT-P_jL\right)+\frac{2\widetilde{\mathfrak{D}}_j^2}{P_j^2}\left(P_jT-E_jL\right)^2, \\
\label{L,T}
\Theta = X_sq_s+X_dq_d,
\qquad
L  =|\boldsymbol{X}_d-\boldsymbol{X}_s|, \qquad T=X_d^0-X_s^0, \\
\label{widetilde_Disp_j}
\widetilde{\mathfrak{D}}_j = \mathfrak{D}_j\left(1+\frac{8ir_jE_\nu^2{\mathfrak{D}}_j^2L}{P_j^3}\right)^{-1/2},
\qquad
\mathfrak{D}_j = \frac{1+\mathfrak{n}r_j}{\sqrt{2R}},
\end{eqnarray}
we arrive at the saddle-point estimate of the function \eref{GeneralizedGreenFunction}:
\begin{eqnarray}
\label{G_j_Final}
\fl\mathbb{G}^j_{\nu\nu'\mu'\mu}
= \Delta_{\nu\nu'}(p_j-p_{\beta})(\hat{p}_j+m_j)\Delta_{\mu'\mu}(p_j+p_{\alpha})
  |\mathbb{V}_d(p_j)\mathbb{V}_s(p_j)|
  \frac{\widetilde{\mathfrak{D}}_je^{-\Omega_j(T,L)-i\Theta}}{i(2\pi)^{3/2}L}.
\end{eqnarray}
This formula can be, and must be, somewhat simplified by putting $r_j=0$ everywhere wherever it
is not a factor multiplying $L$ or $T$ (values which can be arbitrary large).
Then the 4-vector $p_j$ is replaced by $p_\nu=(E_{\nu},\boldsymbol{p}_{\nu})=E_{\nu}l$.
Taking into account the aforesaid, the complex phase \eref{Omega_j} can be written
in an explicitly invariant form
\begin{eqnarray}
\label{Omega_j_inv}
\Omega_j(T,L) = i(p_jX)+\frac{2\widetilde{\mathfrak{D}}_j^2}{E_\nu^2}\left[(p_jX)^2-m_j^2X^2\right],
\qquad
X=X_d-X_s.
\end{eqnarray}
Comparing the factor $\psi_j^*=e^{-\Omega_j}$ in \eref{G_j_Final} with the generic CRGP wavefunction
\eref{psi_AsymptoticExpansion_lowest_star}, we conclude that $\psi_j^*$ can be treated as the (outgoing)
neutrino wave packet in which the role of the parameter $\sigma$ is played by the function
$\Sigma_j=\sqrt{2}\widetilde{\mathfrak{D}}_j/\Gamma_j$ ($\Gamma_j=E_\nu/m_j$).
Since the latter is a complex-valued function, the neutrino wave packet spreads with increase of $L=|\boldsymbol{X}|$.
The spread effect appearing at very large $L$ and leading to both the overall suppression of the amplitude \eref{Amplitude_2}
and modification of the relative contributions with different $j$ into \eref{Amplitude_2}
can be of definite interest for neutrino astrophysics. However, in this paper, we limit ourselves to the analysis
of the probability of the process \eref{Macroprocess_A} under ``terrestrial'' conditions, for which it is
{\it a fortiori} possible to judge that $E_jL\ll\left(\Gamma_jE_j/2\mathfrak{D}_j\right)^2$ and, consequently,
to put $\widetilde{\mathfrak{D}}_j\simeq\mathfrak{D}_j\simeq1/\sqrt{2R}\equiv\mathfrak{D}$ and
$\Sigma_j\simeq\sqrt{2}\mathfrak{D}/\Gamma_j=1/(\Gamma_j\sqrt{R})$.
Hence, the relative energy-momentum uncertainty of the ultrarelativistic neutrino packet,
${\delta}E_j/E_j\sim{\delta}P_j/P_j\sim\mathfrak{D}/E_\nu$, is small and is determined by the momentum
spreads of the external in and out packets.
Of course, the mean position of the neutrino wave packet evolves along the ``classical trajectory''
$\overline{\boldsymbol{L}}=\boldsymbol{v}_jT$, the quantum deviations from which, $\delta\boldsymbol{L}$, are suppressed by the factor
$\exp\left\{-2\mathfrak{D}^2\left[\delta\boldsymbol{L}^2/\Gamma_j^2+\left(\boldsymbol{l}\delta\boldsymbol{L}\right)^2\right]\right\}$.

Now, by applying the identity $P_-\hat{p}_\nu P_+= P_- u_-(\boldsymbol{p}_{\nu})\overline{u}_-(\boldsymbol{p}_{\nu})P_+$, in which
$u_-(\boldsymbol{p}_\nu)$ is the Dirac bispinor for the left-handed massless neutrino and $P_\pm=\frac{1}{2}(1\pm\gamma_5)$,
we define the matrix elements
\begin{eqnarray*}
M_s    = \frac{g^2}{8}\overline{u}_-(\boldsymbol{p}_\nu)\mathcal{J}_s^{\mu}\Delta_{\mu\mu'}(p_\nu+p_{\alpha})O^{\mu'}u(\boldsymbol{p}_\alpha), \cr
M_d^*  = \frac{g^2}{8}\overline{v}(\boldsymbol{p}_\beta)O^{\mu'}\Delta_{\mu'\mu}(p_\nu-p_{\beta})\mathcal{J}_d^{*\mu}u_-(\boldsymbol{p}_\nu),
\end{eqnarray*}
describing, respectively, the production of a \emph{real} massless neutrino $\nu$ in the reaction
$I_s \to F_s'\ell_\alpha^+\nu$ and its absorption in the reaction $\nu I_d \to F_d'\ell_\beta^-$.
Then, taking into account the above-mentioned results, we obtain the final expression for the amplitude \eref{Amplitude_2}:
\begin{eqnarray}
\label{Amplitude_3}
\mathcal{A}_{\beta\alpha}=\frac{\mathfrak{D}|\mathbb{V}_s(p_\nu)\mathbb{V}_d(p_\nu)|M_sM_d^*}{i(2\pi)^{3/2}\mathcal{N}L}
\sum_jV^*_{{\alpha}j}V_{{\beta}j}\,e^{-\Omega_j(T,L)-i\Theta}.
\end{eqnarray}
As is evident from the derivation of this formula and its structure, it is valid for essentially any class of macrodiagrams
with exchange of virtual massive neutrinos between the source and detector vertices, unless we do not specify the explicit form
of the matrix elements $M_s$ and $M_d$. To obtain similar result for the macrodiagrams with exchange of virtual antineutrinos,
one must replace (besides the matrix elements) $\boldsymbol{V}$ by $\boldsymbol{V}^\dag$
(i.e., $V^*_{{\alpha}j} \longmapsto V_{{\alpha}j}$, $V_{{\beta}j} \longmapsto V^*_{{\beta}j}$).

\section{\label{sec:probability} Probability and count rate.}

It can be shown that 
\begin{eqnarray}
\label{SquaredOverlapIntegral}
|\mathbb{V}_{s,d}(p_\nu)|^2 = (2\pi)^4\delta_{s,d}(p_\nu{\mp}q_{s,d})\mathrm{V}_{s,d},
\end{eqnarray}
where $\delta_{s,d}$ (the ``smeared'' $\delta$ functions analogous to the functions $\widetilde{\delta}_{s,d}$)
and $\mathrm{V}_{s,d}$ (the effective four-dimensional overlap volumes of the external packets) are given by
\begin{eqnarray}
\label{delta}
\delta_{s,d}(K) & = (2\pi)^{-2}|\Re_{s,d}|^{-1/2}
\exp\left(-\frac{1}{2}\,\widetilde{\Re}_{s,d}^{\mu\nu}K_{\mu}K_{\nu}\right), \\
\label{mathcalV}
\mathrm{V}_{s,d}
& = \int dx \prod_{\varkappa{\in}S,D}\left|\psi_{\varkappa}\left(\boldsymbol{p}_{\varkappa},x_{\varkappa}-x\right)\right|^2.
\end{eqnarray}
Thus, from \eref{Amplitude_3} we obtain the microscopic probability of the process \eref{Macroprocess_A},
\begin{eqnarray}
\label{MicroscopicProbability_1}
|\mathcal{A}_{\beta\alpha}|^2
 = & \frac{(2\pi)^4\delta_s(p_\nu-q_s)\mathrm{V}_s|M_s|^2}{\prod_{\varkappa{\in}S}2E_{\varkappa}\mathrm{V}_{\varkappa}}\;
     \frac{(2\pi)^4\delta_d(p_\nu+q_d)\mathrm{V}_d|M_d|^2}{\prod_{\varkappa{\in}D}2E_{\varkappa}\mathrm{V}_{\varkappa}} \cr
  & \times\frac{\mathfrak{D}^2}{(2\pi)^3L^2}\Big|\sum_jV^*_{{\alpha}j}V_{{\beta}j}\,e^{-\Omega_j(T,L)}\Big|^2,
\end{eqnarray}
dependent on the parameters $\sigma_\varkappa$, coordinates $x_\varkappa$ and mean momenta $\boldsymbol{p}_\varkappa$
of all packets participated in the reaction. The probability \eref{MicroscopicProbability_1} is small if the product of
the overlap volumes \eref{mathcalV},
\begin{eqnarray*}
\mathrm{V}_s\mathrm{V}_d = (\pi/2)^4\left(|\Re_s||\Re_d|\right)^{-1/2}\exp\left[-2\left(\mathfrak{S}_s+\mathfrak{S}_d\right)\right],
\end{eqnarray*}
vanishes, i.e.\ if the external wave packets in the source and detector do not overlap in the spacetime regions
surrounding the impact points $X_s$ and $X_d$.

Let us note that the 4-vector $p_\nu$ is a function of $\boldsymbol{p}_\varkappa$ and $\sigma_\varkappa$, and
$p_\nu=q_s=-q_d$ in the plane-wave limit ($\sigma_\varkappa=0$, $\forall\varkappa$).
Thus, for sufficiently small $\sigma_\varkappa$,
\begin{eqnarray*}
\delta_s(p_\nu-q_s)\delta_d(p_\nu+q_d)\approx\delta_s(0)\delta_d(0)=(2\pi)^{-4}\left(|\Re_s||\Re_d|\right)^{-1/2}.
\end{eqnarray*}
But what controls the approximate equality of the 4-momentum transfers $q_s$ and $-q_d$? To answer this question,
it is useful to rewrite \eref{MicroscopicProbability_1} in the fashion used by Cardall for a similar purpose~\cite{Cardall:99b}.
By using the explicit form of the functions $\delta_{s,d}$ and $\mathfrak{D}$, one can prove the following approximate relation:
\begin{eqnarray}
\label{Cardal'sTrick}
\fl 2\sqrt{\pi}\mathfrak{D}\delta_s\left(p_\nu-q_s\right)\delta_d\left(p_\nu+q_d\right)F(p_\nu)
 = \int dE'_{\nu}\delta_s\left(p'_{\nu}-q_s\right)\delta_d\left(p'_{\nu}+q_d\right)F(p'_{\nu}),
\end{eqnarray}
which is valid with the same accuracy with which the amplitude \eref{Amplitude_3} itself has been deduced
(that is, with the accuracy of the saddle-point method).
Here $F(p_\nu)$ is an arbitrary slowly varying function and $p'_{\nu}=(E'_{\nu},\boldsymbol{p}'_{\nu})=E'_{\nu}l$.
With the help of the relation \eref{Cardal'sTrick} the microscopic probability \eref{MicroscopicProbability_1} transforms to
\begin{eqnarray}
\label{MicroscopicProbability_2}
|\mathcal{A}_{\beta\alpha}|^2
= & \int dE_{\nu}\,\frac{(2\pi)^4\delta_s(p_\nu-q_s)\mathrm{V}_s|M_s|^2}{\prod_{\varkappa{\in}S}2E_{\varkappa}\mathrm{V}_{\varkappa}}\;
    \frac{(2\pi)^4\delta_d(p_\nu+q_d)\mathrm{V}_d|M_d|^2}{\prod_{\varkappa{\in}D}2E_{\varkappa}\mathrm{V}_{\varkappa}} \cr
  & \times\frac{\mathfrak{D}}{2\sqrt{\pi}(2\pi)^3L^2}\Big|\sum_jV^*_{{\alpha}j}V_{{\beta}j}\,e^{-\Omega_j(T,L)}\Big|^2,
\end{eqnarray}
where we have omitted the prime on the integration variable $E_\nu$, but now $E_\nu$
(as well as  $\boldsymbol{p}_{\nu}=E_{\nu}\boldsymbol{l}$) is in no way related to the parameters
of the external wave packets.
Expressions \eref{MicroscopicProbability_1} and \eref{MicroscopicProbability_2} are equivalent within the adopted accuracy,
but from \eref{MicroscopicProbability_2} it is apparent that the energy-momentum conservation is governed by the
factors $\delta_s(p_\nu-q_s)$ and $\delta_d(p_\nu+q_d)$ which, at sufficiently small $\sigma_\varkappa$,
could be substituted by the usual $\delta$ functions.

The probability \eref{MicroscopicProbability_2} is the most general result of this paper.
However, it is \emph{too} general to be directly applied to the contemporary neutrino oscillation experiments.
In order to obtain the actually observable quantities, the probability \eref{MicroscopicProbability_2} should be properly
averaged/integrated over all the unmeasurable or unused variables of incoming/outgoing wave-packet states.
Let us call this procedure the macroscopic averaging.
Clearly, such a procedure can only be realized by taking into account the physical conditions
of the real experimental environment.
For these reasons and in this sense, further analysis becomes model-dependent.
 
As a simple but realistic example, we consider a thought experiment in which it is assumed that the
statistical distributions of the incoming packets $a \in I_{s,d}$ over the mean momenta, spin projections
and spacetime coordinates in the source and detector ``devices'' can be described by the one-particle
distribution functions $f_a(\boldsymbol{p}_a,s_a,x_a)$. It is convenient to normalize each function $f_a$
to the total number, $N_a(x_a^0)$, of the packets $a$ at a time $x_a^0$:
\begin{eqnarray}
\label{f_normalization}
\sum\limits_{s_a}\int\frac{d\boldsymbol{x}_a d\boldsymbol{p}_a}{(2\pi)^3}f_a(\boldsymbol{p}_a,s_a,x_a) = N_a(x_a^0)
\qquad
(a \in I_{s,d}).
\end{eqnarray}
For clarity purposes, we must define (or rather redefine) the terms ``source'' and ``detector'' which were
so far used for designating the blocks of the macrodiagrams.
In what follows we will use these terms and notation $\mathcal{S}$ and $\mathcal{D}$ both for the corresponding devices
and, more abstractly, for the supports of the products of the distribution functions $f_a$ in the spacetime variables
(namely, $\mathcal{S}=\mathrm{supp}_{\{x_a\}}\prod_{a}f_a$, $a{\in}I_s$ and similarly for $\mathcal{D}$),
which are assumed to be finite and mutually disjoint within the space domain. 
We further suppose that the effective spatial dimensions of $\mathcal{S}$ and $\mathcal{D}$ are small compared
to the mean distance between them but very large compared to the effective dimensions ($\sim\sigma_\varkappa^{-1}$)
of all wave packets moving inside $\mathcal{S}$ and $\mathcal{D}$. For definiteness, we also approve that
the experiment measures only the momenta of the secondaries in $\mathcal{D}$, and (owing to the large distance between
$\mathcal{S}$ and $\mathcal{D}$) the background events caused by the secondaries falling into $\mathcal{D}$ from $\mathcal{S}$
can be neglected. 
Lastly, we accept the detection efficiency to be 100\%, though the formalism allows a straightforward account
for the real efficiency, acceptance, etc.
With all these assumptions, the macroscopically averaged probability \eref{MicroscopicProbability_2} reads
\begin{eqnarray}
\label{AveragedProbability_1}
\fl\langle\!\langle|\mathcal{A}_{\beta\alpha}|^2\rangle\!\rangle
= & \sum\limits_{\mathrm{spins}}
          \int\prod_{a{\in}I_s}\frac{d\boldsymbol{x}_a d\boldsymbol{p}_af_a(\boldsymbol{p}_a,s_a,x_a)}{(2\pi)^32E_a\mathrm{V}_a}
          \int\prod_{b{\in}F_s}\frac{d\boldsymbol{x}_b d\boldsymbol{p}_b}{(2\pi)^32E_b\mathrm{V}_b}\mathrm{V}_s   \cr
  & \times\int \prod_{a{\in}I_d}\frac{d\boldsymbol{x}_a d\boldsymbol{p}_af_a(\boldsymbol{p}_a,s_a,x_a)}{(2\pi)^3 2E_a\mathrm{V}_a}
          \int\prod_{b{\in}F_d}\frac{d\boldsymbol{x}_b [d\boldsymbol{p}_b]}{(2\pi)^32E_b\mathrm{V}_b}\mathrm{V}_d \cr
  & \times\int dE_{\nu}(2\pi)^4{\delta}_s(p_\nu-q_s)|M_s|^2 (2\pi)^4{\delta}_d(p_\nu+q_d)|M_d|^2                  \cr
  & \times\frac{\mathfrak{D}}{2\sqrt{\pi}(2\pi)^3 L^2}\Big|\sum\limits_j V^*_{{\alpha}j}V_{{\beta}j}\,e^{-\Omega_j(T,L)}\Big|^2.
\end{eqnarray}
Here and below the symbol $\sum_{\mathrm{spins}}$ denotes the averaging over the spin projections of the initial
states and summation over the spin projections of the final states in $\mathcal{S}$ and $\mathcal{D}$.
The symbol $[d\boldsymbol{p}_b]$ indicates that integration in the variable $\boldsymbol{p}_b$ is not performed, i.e.\
$\int[d\boldsymbol{p}_b]=d\boldsymbol{p}_b$.
With regard to the normalization conditions \eref{f_normalization}, it is easy to recognise that \eref{AveragedProbability_1}
represents the total number, $dN_{\alpha\beta}$, of events recorded in $\mathcal{D}$ and consisted of the secondary wave packets
$b{\in}F_d$ having the mean momenta between $\boldsymbol{p}_b$ and $\boldsymbol{p}_b+d\boldsymbol{p}_b$.
Under additional assumptions, the somewhat unwieldy expression \eref{AveragedProbability_1} can be simplified in a few steps.

An approximate multidimensional integration over the spatial variables in \eref{AveragedProbability_1} can be performed using
the integral representation for the overlap volumes \eref{mathcalV} and taking into account that the distribution
functions $f_a$, as well as the factors $e^{-\Omega_j(T,L)-\Omega_i^*(T,L)}/L^2$ are \emph{assumed} to vary at large
(macroscopic) scales, whereas the integrand
$\prod_{\varkappa}\left|\psi_{\varkappa}\left(\boldsymbol{p}_{\varkappa},x_{\varkappa}-x\right)\right|^2$ 
in \eref{mathcalV} is essentially different from zero only if the classical word lines of all packets $\varkappa$
pass through a small (not necessarily microscopic) vicinity of the integration variable
(let the latter be $x$ for $\mathrm{V}_s$ and $y$ for $\mathrm{V}_d$).
Hence, neglecting the edge effects, it is safe to replace $x_{\varkappa}$ by $x$ ($y$) for $\varkappa \in S$ ($D$)
in the mentioned slowly varying factors. Then, as is seen from \eref{X_sd}, $X_s=x$ and $X_d=y$.
The remaining integrals over the variables $\boldsymbol{x}_\varkappa$ yield the factor
$\prod_{\varkappa{\in}S{\oplus}D}\mathrm{V}_\varkappa$,
which cancels the same factor in the denominator of the integrand in \eref{AveragedProbability_1}.
As a result, we can rewrite \eref{AveragedProbability_1} as follows:
\begin{eqnarray}
\label{AveragedProbability_2}
\fl dN_{\alpha\beta}
= \sum\limits_{\mathrm{spins}} \int dx \int dy\,\int d\mathfrak{P}_s \int d\mathfrak{P}_d \int dE_{\nu}
  \frac{\mathfrak{D}\,\left|\sum_j V^*_{{\alpha}j}V_{{\beta}j}\,
  e^{-\Omega_j(T,L)}\right|^2}{16\pi^{7/2}|\boldsymbol{y}-\boldsymbol{x}|^2},
\end{eqnarray}
where we have defined the differential forms 
\begin{eqnarray}
\label{Differential_s}
\fl d\mathfrak{P}_s & = \prod_{a{\in}I_s}\frac{d\boldsymbol{p}_af_a(\boldsymbol{p}_a,s_a,x)}{(2\pi)^32E_a}\prod_{b{\in}F_s}
                    \frac{d \boldsymbol{p}_b }{(2\pi)^32E_b}(2\pi)^4{\delta}_s(p_\nu-q_s)|M_s|^2, \\
\label{Differential_d}
\fl d\mathfrak{P}_d & = \prod_{a{\in}I_d}\frac{d\boldsymbol{p}_af_a(\boldsymbol{p}_a,s_a,y)}{(2\pi)^32E_a}\prod_{b{\in}F_d}
                    \frac{[d\boldsymbol{p}_b]}{(2\pi)^32E_b}(2\pi)^4{\delta}_d(p_\nu+q_d)|M_d|^2.
\end{eqnarray}
The phase $\Omega_j(T,L)$ in \eref{AveragedProbability_2} is still defined by equation \eref{Omega_j} or \eref{Omega_j_inv},
in which, now $T=X_0=y_0-x_0$ and $L=|\boldsymbol{X}|=|\boldsymbol{y}-\boldsymbol{x}|$.
It must be underlined that the considerations used for deriving \eref{AveragedProbability_2} were based on the
additional restrictions, which may not be fully adequate to the particular experimental conditions.%
\footnote{Moreover, we implicitly used the current experimental constraints on the neutrino masses, which suggest
          that $\mathrm{Im}(\Omega_j+\Omega_i^*)$ noticeably vary on the macroscopically large scales
          $L_{ij} \propto E_\nu/|m_i^2-m_j^2|$ and
          $\mathrm{Re}(\Omega_j+\Omega_i^*)$ -- on the scales much larger than $L_{ij}$.
          We have to remind, however, that such conclusions are based on the QM (rather than QFT)
          analyses of the existing data.}
Thus the comparatively simple but approximate formula \eref{AveragedProbability_2} is not entirely equivalent
to the more sophisticated and general result \eref{AveragedProbability_1}.

For further simplification, we perform in \eref{AveragedProbability_2} integration in time variables $x^0$ and $y^0$.
This integration can be easily done in assumption that, during the experiment, the distribution functions $f_a$ in $\mathcal{S}$
and $\mathcal{D}$ vary slowly enough with time so that they can be modelled by the ``rectangular ledges'' 
\begin{eqnarray}
\label{Steps}
\begin{array}{ll}
f_a(\boldsymbol{p}_a,s_a;x) = \theta\left(x^0-x^0_1\right)\theta\left(x^0_2-x^0\right)\overline{f}_a(\boldsymbol{p}_a,s_a;\boldsymbol{x})
\enskip\mathrm{for}\enskip  a{\in}I_{s}, \cr
f_a(\boldsymbol{p}_a,s_a;y) = \theta\left(y^0-y^0_1\right)\theta\left(y^0_2-y^0\right)\overline{f}_a(\boldsymbol{p}_a,s_a;\boldsymbol{y})
\enskip\mathrm{for}\enskip  a{\in}I_{d}.
\end{array}
\end{eqnarray}
In case of detector, the step functions in \eref{Steps} can be thought as the ``hardware'' or ``software'' trigger conditions.
The periods of stationarity $\tau_s=x^0_2-x^0_1$ and $\tau_d=y^0_2-y^0_1$ can be astronomically long, as it is for the solar
and atmospheric neutrino experiments ($\tau_s \ggg \tau_d$ in these cases), or very short, like in the experiments with short-pulsed
accelerator beams (when usually $\tau_s\lesssim\tau_d$), but we anyhow assume that the time intervals needed to switch on and switch off
the source (detector) are negligibly small in comparison with $\tau_s$ ($\tau_d$).
Within the model \eref{Steps}, the only time-dependent factor in the integrand of \eref{AveragedProbability_2}
is $e^{-\Omega_j-\Omega_i^*}$. So the problem is reduced to the simple integral
\begin{eqnarray}
\label{Int_x_0_y_0_final}
\fl\int_{y^0_1}^{y^0_2}dy^0\,\int_{x^0_1}^{x^0_2}dx^0\;e^{-\Omega_j(y^0-x^0,L)-\Omega_i^*(y^0-x^0,L)}
= \frac{\sqrt{\pi}}{2\mathfrak{D}}\tau_d\exp\left(i\varphi_{ij}-\mathscr{A}_{ij}^2\right)S_{ij}.
\end{eqnarray}
In this relation, we have adopted the following notation:
\begin{eqnarray}
\label{S_ij}
S_{ij}           = \frac{\exp\left(-\mathscr{B}_{ij}^2\right)}{4\tau_d\mathfrak{D}}\sum_{l,l'=1}^2(-1)^{l+l'+1}
                    \mathrm{Ierf}\left[2\mathfrak{D}\left(x^0_{l}-y^0_{l'}+\frac{L}{v_{ij}}\right)-i\mathscr{B}_{ij}\right], \\
\label{PhaseTerms}
\mathscr{A}_{ij} = (v_j-v_i){\mathfrak{D}}L
                 = \frac{2\pi{\mathfrak{D}}L}{E_{\nu}L_{ij}},   \qquad
\mathscr{B}_{ij} = \frac{{\Delta}E_{ji}}{4\mathfrak{D}}
                 = \frac{{\pi}\mathfrak{n}}{2{\mathfrak{D}}L_{ij}},
\end{eqnarray}
where
\begin{eqnarray*}
\varphi_{ij}     = \frac{2\pi L}{L_{ij}},                       \qquad
L_{ij}           = \frac{4{\pi}E_{\nu}}{{\Delta}m_{ij}^2},      \qquad
\frac{1}{v_{ij}} = \frac{1}{2}\left(\frac{1}{v_i}+\frac{1}{v_j}\right),
\end{eqnarray*}
\begin{eqnarray*}
{\Delta}m_{ij}^2 = m_i^2-m_j^2,
\qquad
{\Delta}E_{ij}   = E_i-E_j,
\end{eqnarray*}
\begin{eqnarray*}
\label{Ierf}
\mathrm{Ierf}(z) = \int_0^zdz'\mathrm{erf}(z')+\frac{1}{\sqrt{\pi}}=z\,\mathrm{erf}(z)+\frac{1}{\sqrt{\pi}}e^{-z^2}, 
\end{eqnarray*}
and $\mathrm{erf}(z)$ is the error function.
For a more realistic description of the accelerator beam pulse experiments, the model \eref{Steps} could be readily extended
by inclusion of a series of rectangular ledges followed by pauses during which $f_a=0$. 
We will, however, proceed with the simplest case, which reproduces the most significant effects. 
Then substituting  \eref{Int_x_0_y_0_final} into \eref{AveragedProbability_2} we obtain:
\begin{eqnarray}
\label{AveragedProbability_3_a}
dN_{\alpha\beta}
& =  \tau_d\sum\limits_{\mathrm{spins}}\int d\boldsymbol{x}\int d\boldsymbol{y}\int d\mathfrak{P}_s\int d\mathfrak{P}_d\int dE_{\nu} 
  \frac{\mathcal{P}_{\alpha\beta}(E_\nu,|\boldsymbol{y}-\boldsymbol{x}|)}{4(2\pi)^3|\boldsymbol{y}-\boldsymbol{x}|^2} \\
\label{AveragedProbability_3_b}
& = \frac{\tau_d}{V_{\mathcal{D}}V_{\mathcal{S}}}\int d\boldsymbol{x}\int d\boldsymbol{y}\int d\Phi_\nu\int
     d\sigma_{{\nu}\mathcal{D}}\mathcal{P}_{\alpha\beta}(E_\nu,|\boldsymbol{y}-\boldsymbol{x}|).
\end{eqnarray}
The differential forms $d\mathfrak{P}_{s,d}$ in \eref{AveragedProbability_3_a} are defined according to equations
\eref{Differential_s} and \eref{Differential_d} in which the distribution functions $f_a$ should be substituted by $\overline{f}_a$.
Equation~\eref{AveragedProbability_3_b} is written by applying the identity
\begin{eqnarray*}
\sum\limits_{\mathrm{spins}}\frac{d\mathfrak{P}_sd\mathfrak{P}_ddE_\nu}{4(2\pi)^3|\boldsymbol{y}-\boldsymbol{x}|^2}
& = \sum\limits_{\mathrm{spins}\,{\in}\,S}\frac{d\mathfrak{P}_sd\boldsymbol{p}_\nu}
  {(2\pi)^32E_{\nu}|\boldsymbol{y}-\boldsymbol{x}|^2d\boldsymbol{\Omega}_{\nu}}
  \sum\limits_{\mathrm{spins}\,{\in}\,D}\frac{d\mathfrak{P}_d}{2E_{\nu}} \cr
& \equiv \frac{d\Phi_{\nu}\,d\sigma_{{\nu}\mathcal{D}}}{V_{\mathcal{D}}V_{\mathcal{S}}},
\end{eqnarray*}
where $V_{\mathcal{S}}$ and $V_{\mathcal{D}}$ are the spatial volumes of the source and detector, respectively.
The differential form $d\Phi_\nu$ is defined in such a way that the integral
\begin{eqnarray}
\label{NeutrinoFlux}
\frac{d\boldsymbol{x}}{V_{\mathcal{S}}}\int \frac{d\Phi_\nu}{dE_{\nu}}
=  d\boldsymbol{x}\sum\limits_{\mathrm{spins}\,{\in}\,S}\int\frac{d\mathfrak{P}_sE_{\nu}}
  {2(2\pi)^3|\boldsymbol{y}-\boldsymbol{x}|^2}
\end{eqnarray}
is nothing else than the flux density of neutrinos in $\mathcal{D}$, produced through the processes
$I_s \to F_s'\ell_\alpha^+\nu$ in $\mathcal{S}$. More precisely, it  is the number of neutrinos appearing per unit time
and unit neutrino energy in an elementary volume $d\boldsymbol{x}$ around the point $\boldsymbol{x}\in\mathcal{S}$,
travelling within the solid angle $d\boldsymbol{\Omega}_{\nu}$ about the flow direction
$\boldsymbol{l}=(\boldsymbol{y}-\boldsymbol{x})/|\boldsymbol{y}-\boldsymbol{x}|$
and crossing a unit area, placed around the point $\boldsymbol{y}\in\mathcal{D}$ and normal to $\boldsymbol{l}$.
The quantity $d\sigma_{{\nu}\mathcal{D}}$ is defined in such a way that
\begin{eqnarray}
\label{CrossSection}
\frac{1}{V_{\mathcal{D}}}\int d\boldsymbol{y}d\sigma_{{\nu}\mathcal{D}}
= \sum\limits_{\mathrm{spins}\,{\in}\,D}\int \frac{d\boldsymbol{y} d\mathfrak{P}_d}{2E_{\nu}}
\end{eqnarray}
represents the differential cross section of the neutrino scattering off the detector as a whole.
In the particular (and the most basically important) case of neutrino scattering in the reaction $\nu a \to F_d'\ell_\beta^-$,
provided that the momentum distribution of the target scatterers $a$ is sufficiently narrow, the differential form
$d\sigma_{{\nu}\mathcal{D}}$ becomes exactly the elementary differential cross section
of this reaction multiplied by the total number of the particles $a$ in $\mathcal{D}$.

Now let us address the last subintegral multiplier of \eref{AveragedProbability_3_b}, given by
\begin{eqnarray}
\label{TransitionProbability}
\mathcal{P}_{\alpha\beta}(E_\nu,L) = \sum_{ij}V^*_{{\alpha}i}V_{{\alpha}j}V^*_{{\beta}j}V_{{\beta}i}
S_{ij}\exp\left(i\varphi_{ij}-\mathscr{A}_{ij}^2\right).
\end{eqnarray}
This factor coincides with the well-known QM expression for the neutrino flavor transition probability,
provided that $S_{ij}=1$ and $\mathscr{A}_{ij}=0$, and thus it can be considered as a QFT refinement of the QM result. 
However a probabilistic interpretation of the function $\mathcal{P}_{\alpha\beta}$ can be only provisionally true,
because the factors $S_{ij}$ and $\mathscr{A}_{ij}$ in \eref{TransitionProbability} involve the functions
$\mathfrak{D}$ and $\mathfrak{n}$ strongly dependent on the neutrino energy $E_\nu$ and external momenta $\boldsymbol{p}_{\varkappa}$;
all these (except for the momenta of secondaries in $\mathcal{D}$) are variables of integration in \eref{AveragedProbability_3_b}.
As a result, the factor $\mathcal{P}_{\alpha\beta}$, as function of $\alpha$ and $\beta$, does not satisfy the unitarity relations
\[\sum_{\beta}\mathcal{P}_{\alpha\beta}=\sum_{\alpha}\mathcal{P}_{\alpha\beta}=1,\] which are a commonplace in the QM theory.
The point is that the domains and shapes of the functions $\mathfrak{D}$ and $\mathfrak{n}$ are essentially
different for each of the nine leptonic pairs $(\ell_\alpha,\ell_\beta)$.
These differences are governed by the kinematics of the subprocesses in $\mathcal{S}$ and $\mathcal{D}$
(in particular, their thresholds), that is, eventually, by the leptonic masses ($m_e$, $m_\mu$, $m_\tau$)
and by the momentum spreads ($\sigma_e$, $\sigma_\mu$, $\sigma_\tau$) of the leptonic wave packets,
which are not necessarily equal to each other, perhaps even within an order of magnitude.
The probabilistic treatment of $\mathcal{P}_{\alpha\beta}$ is even more problematic in real-life experiments,
because the detector event rate (with $\ell_\beta$ appearance in our case) is defined by many subprocesses
of different types in the source and detector.
For example, in the atmospheric and accelerator neutrino experiments, the major processes of neutrino
          production are in-flight decays of light mesons ($\pi_{\mu2}$, $K_{\mu2}$, $K_{\mu3}$, $K_{e3}$, etc) and muons,
          and neutrino interactions with detector consist of an incoherent superposition of exclusive reactions of many types,
          -- from (quasi)elastic to deep inelastic.

One more technical drawback to interpretation of \eref{TransitionProbability} is the dependence of the function $S_{ij}$
(which will be referred to as a decoherence factor) on the four ``instrumental'' time parameters $x^0_1$, $x^0_2$, $y^0_1$
and $y^0_2$. So far we have made no assumption concerning
a ``synchronization'' of the time windows $(x^0_1,x^0_2)$ and $(y^0_1,y^0_2)$.
Thus, it is no wonder that the decoherence factor turns to be vanishingly small in magnitude if these windows
are not adjusted to account that the representative time of ultrarelativistic neutrino propagation from
$\mathcal{S}$ to $\mathcal{D}$ is equal to the mean distance, $\overline{L}$, between $\mathcal{S}$ and $\mathcal{D}$.
Before discussing the role of the decoherence factor, we perform one more, and the last, simplification of the formula
for $dN_{\alpha\beta}$, again using the requirement that the characteristic dimensions of $\mathcal{S}$ and $\mathcal{D}$
are small compared to $\overline{L}$.
Under certain conditions, this allows us to replace $|\boldsymbol{y}-\boldsymbol{x}|$
in the integrand of \eref{AveragedProbability_3_b} by $\overline{L}$ and the differential forms $d\Phi_\nu$ and
$d\sigma_{{\nu}\mathcal{D}}$ by their averages, $d\overline{\Phi}_\nu$ and $d\overline{\sigma}_{{\nu}\mathcal{D}}$,
over the spatial volumes of $\mathcal{S}$ and $\mathcal{D}$, respectively.
Finally, we arrive at the approximate formula
\begin{eqnarray}
\label{AveragedProbability_4}
dN_{\alpha\beta}
= \tau_d\int d\overline{\Phi}_\nu\int d\overline{\sigma}_{{\nu}\mathcal{D}}
  \mathcal{P}_{\alpha\beta}(E_\nu,\overline{L}).
\end{eqnarray}
Its range of applicability is in general much more limited than that of \eref{AveragedProbability_3_b},
as a consequence of additional restrictions implicitly imposed on the distribution functions $\overline{f}_a$,
absolute dimensions and geometry of $\mathcal{S}$ and $\mathcal{D}$. These issues are not discussed in this paper
but must be the subject of special attention in the neutrino oscillation experiments.

Let us now return to the decoherence factor, limiting ourselves to a consideration of ``synchronized'' measurements,
in which $x^0_{1,2}=\mp\tau_s/2$, $y^0_{1,2}=\overline{L}{\mp}\tau_d/2$.
With certain technical simplifications, the factor \eref{S_ij} can be expressed through  
a real-valued function $S(t,t',b)$ of three dimensionless variables, namely
\begin{eqnarray*}
S_{ij}=S\left(\mathfrak{D}\tau_s,\mathfrak{D}\tau_d,\mathscr{B}_{ij}\right),
\end{eqnarray*}
\begin{eqnarray*}
2t'S(t,t',b) = \exp\left(-b^2\right)\mathrm{Re}\left[\mathrm{Ierf}\left(t+t'+ib\right)-\mathrm{Ierf}\left(t-t'+ib\right)\right].
\end{eqnarray*}
Figure~\ref{fig:S_0(t,t')} shows the numerically evaluated 3D plot and 2D density plot of the function $S_0(t,t')=S(t,t',0)$.
\begin{figure}[htb]
\centering
~\includegraphics[height=0.45\linewidth]{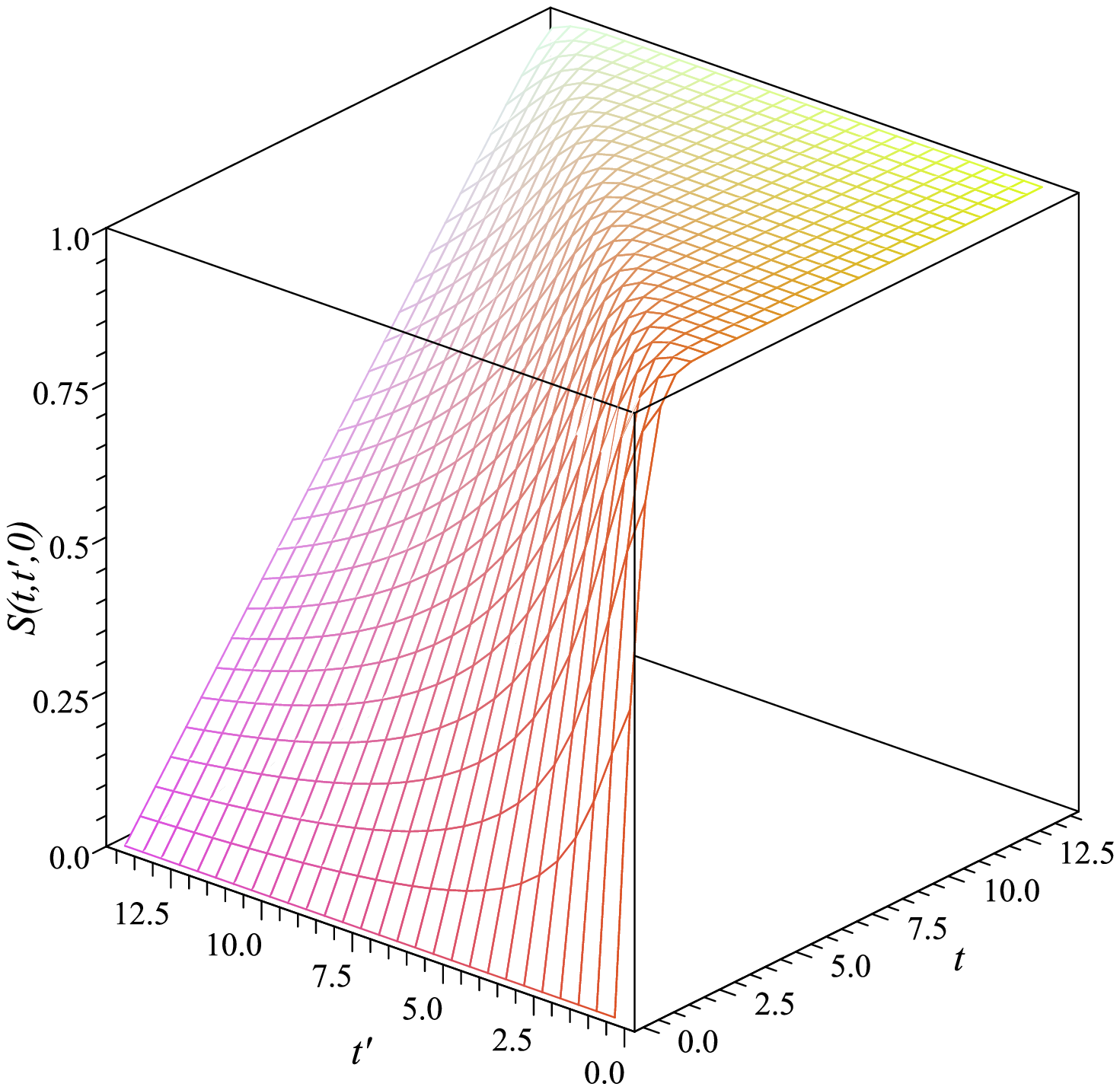}
\hfill
 \includegraphics[height=0.45\linewidth]{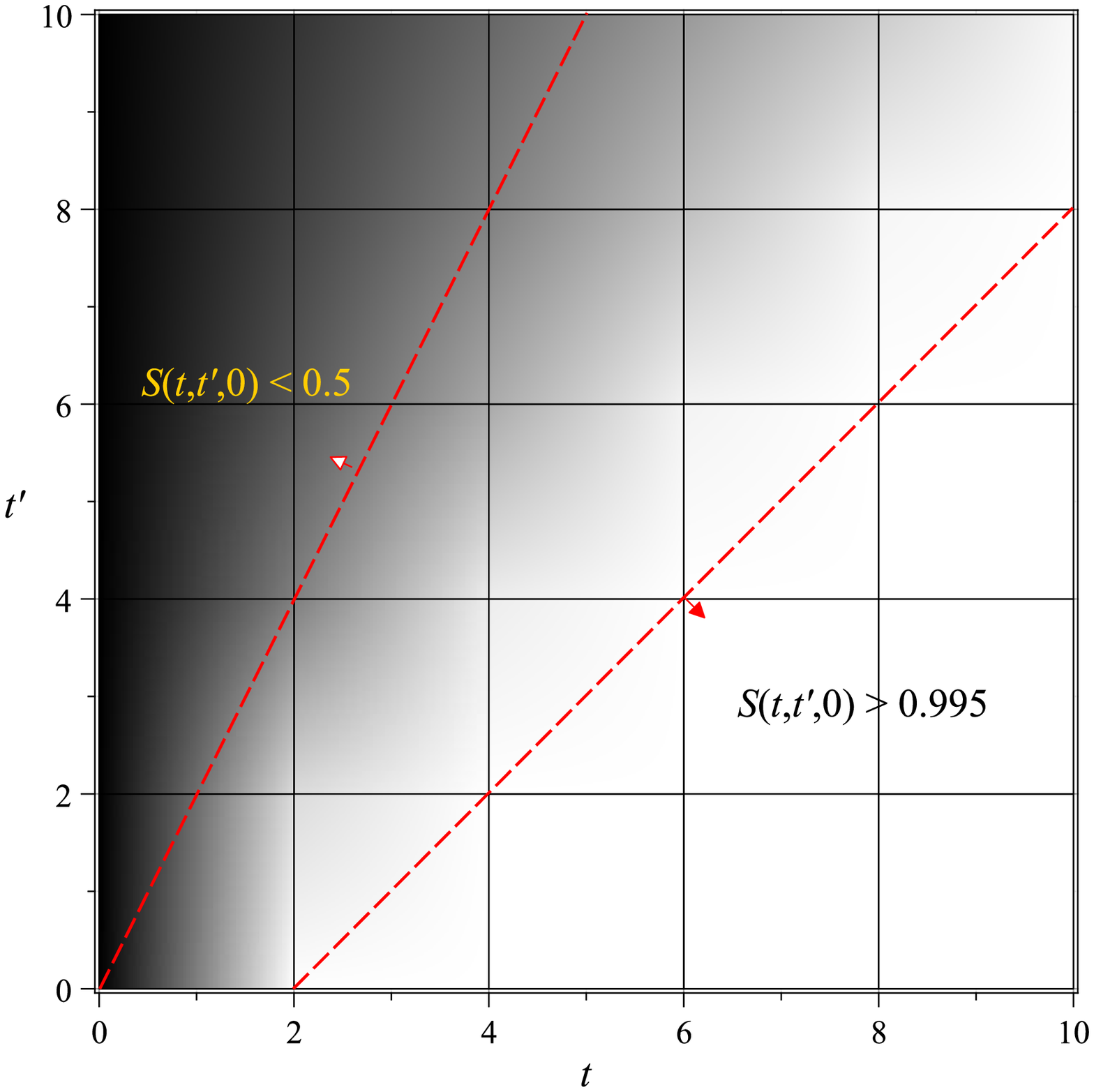}~
\protect\caption[A 3D plot (left) and a 2D density plot (right) of the decoherence function $S_0(t,t')$.]
                {A 3D plot (left) and a 2D density plot (right) of the decoherence function $S_0(t,t')$.
                 The darker regions in the 2D plot correspond to the smaller values of $S_0(t,t')$.}
\label{fig:S_0(t,t')}
\end{figure}
It can be proved that $0<S_0(t,t')<1$ for any $t,t'>0$ and $S_0(t,t')<t/t'$ for $t' \ge t$. This implies that the mean
count rate in the detector, $dR_{\alpha\beta}=dN_{\alpha\beta}/\tau_d$, decreases with a rise of the ratio $\tau_d/\tau_s>1$.
The reason is apparent: the number of events recorded in $\mathcal{D}$ cannot be larger then the number of neutrinos
emitted from $\mathcal{S}$.
A less obvious conclusion follows from the inequality $S_0(t+{\delta}t,t)>\mathrm{erf}({\delta}t)$ valid for ${\delta}t>0$:
the count rate is not suppressed at sufficiently large $\tau_s/\tau_d$.
Clearly, this condition is over-fulfilled in the solar and atmospheric neutrino experiments but may be violated
in the accelerator and perhaps reactor experiments.
Two dashed lines in the right panel of figure~\ref{fig:S_0(t,t')} separate the regions in which $S_0<0.5$ ($\tau_s<2\tau_d$) and $S_0>0.995$
($\tau_s>\tau_d+2/\mathfrak{D}$).

In the particular case $t=t'$, which is of interest for accelerator experiments, the function $S_0(t,t)$ converges
to unity only at very large $t$ (in practice for $t\gtrsim100$).
Thus, in order to set $S_0(t,t')=1$, the detector exposition time $\tau_d$ must be either sufficiently small in comparison with
$\tau_s$ or (at $\tau_d\approx\tau_s$) much larger than the characteristic time scale $\tau_\nu = 1/\min(\mathfrak{D})$,
where the minimum should be taken over the phase subspaces of the process \eref{Macroprocess_A}, which are responsible
for a significant contribution into the count rate.

On the other hand, the observed strong dependence of the common suppression factor $S_0(t,t')$ on its arguments at $t \lesssim t'$
provides a potential possibility of an experimental estimation of the function $\mathfrak{D}$ (or, rather, of its mean values within
the above-mentioned phase subspaces), based on the measuring the count rate $dR_{\alpha\beta}$ as a function of $\tau_d$ and $\tau_s$
(at fixed $\overline{L}$), and comparing the data with the results of Monte-Carlo simulations. The optimal strategy of such an experiment
should be a subject of a dedicated analysis.

The profiles of the function $S(t,t',b)$ calculated at various fixed values of $b$ are displayed in figure~\ref{fig:S(t,t',b)}.
It appears that the shape of $S(t,t',b)$ as a function of $t$ and $t'$ becomes more and more complicated with an increase
of $b$. For $b>3-4$, $S(t,t',b)$ rapidly oscillates around zero, highly suppressing the ``non-diagonal'' (with $i \ne j$)
terms in the factor $\mathcal{P}_{\alpha\beta}(E_\nu,\overline{L})$.
\begin{figure}[htb]
\centering
\includegraphics[width=\linewidth]{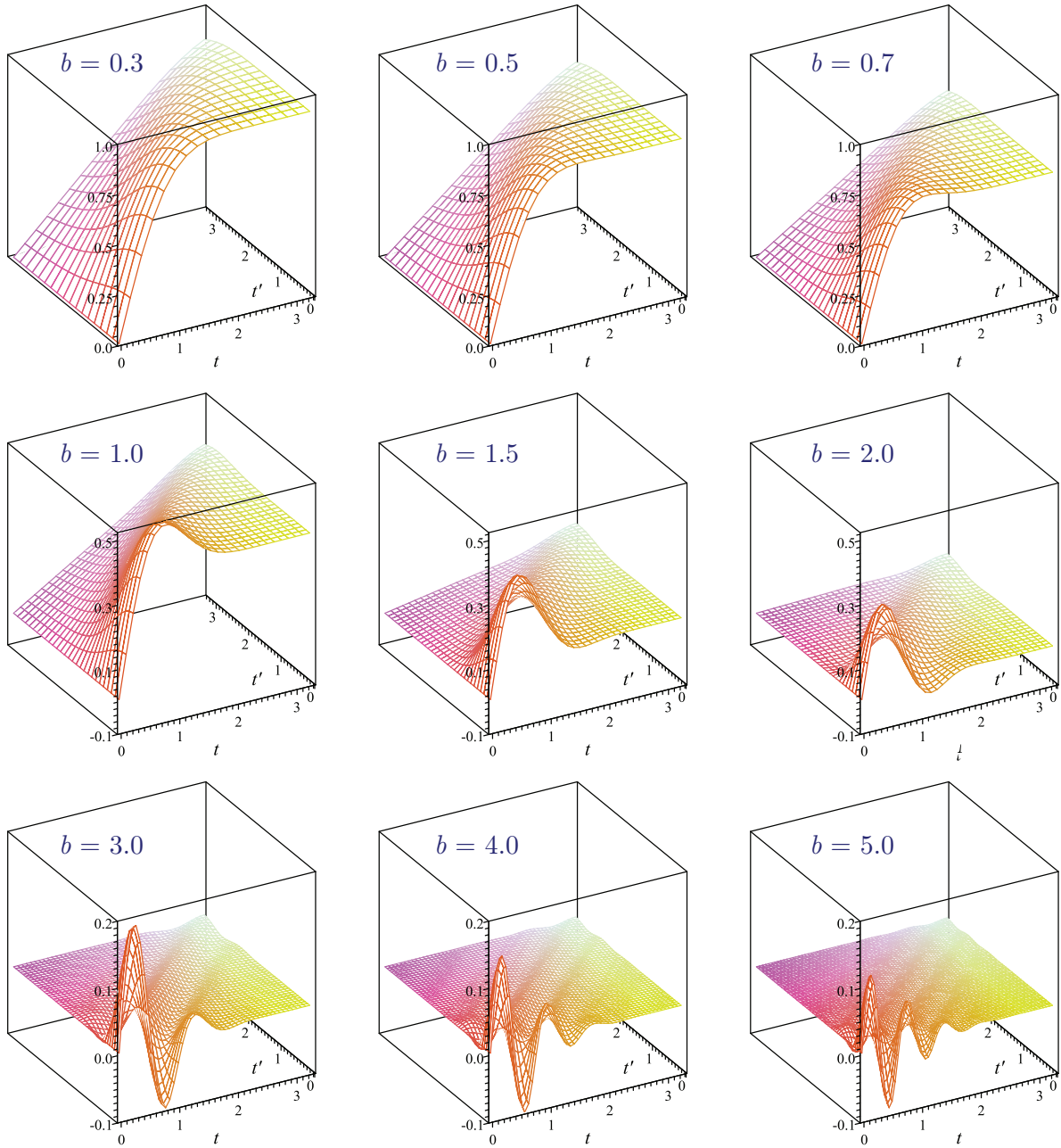}
\protect\caption{Profiles of the decoherence function $S(t,t',b)$ calculated at nine values of $b$ shown on the panels.
}
\label{fig:S(t,t',b)}
\end{figure}
Figure~\ref{fig:S(t,t,b)} shows the dependence of the function $S(t,t,b)$ on the parameter $b$, evaluated at various fixed
values of $t$.
At large $t$, this dependence has a quasi-periodic character superimposed by a fast decrease of $S(t,t,b)$ with increasing $b$.
At very large $t$, the function $S(t,t,b)$ becomes nearly independent of $t$, slowly approaching the asymptotic behavior
\[
S(t,t',b)\sim\exp(-b^2) \qquad (t,t'\to\infty).
\]
\begin{figure}[htb]
\centering
\includegraphics[width=\linewidth]{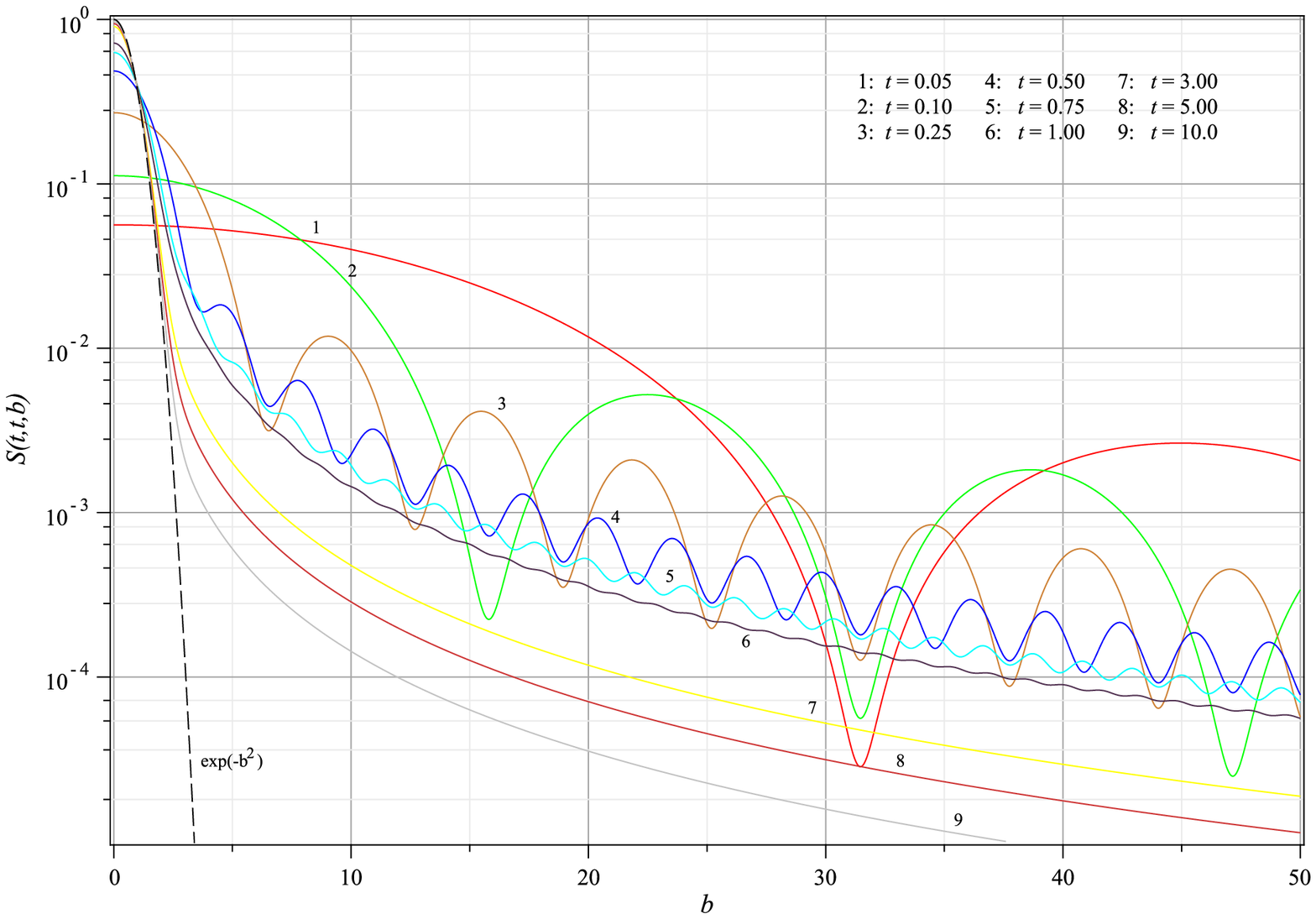}
\protect\caption{The function $S(t,t,b)$ versus $b$ at fixed values of $t$ shown in legend.
                 The asymptotics of  $S(t,t,b)$ at $t\to\infty$ and finite $b$ is shown for comparison.}
\label{fig:S(t,t,b)}
\end{figure}
In this asymptotic regime, the ``probability'' \eref{TransitionProbability} takes on the form already known from the literature
(see, e.g., \cite{Giunti-Kim:07,Beuthe} and references therein),
\begin{eqnarray}
\label{TransitionProbability_limit} 
\mathcal{P}_{\alpha\beta}(E_\nu,\overline{L})
=  \sum_{ij} V^*_{{\alpha}i}V_{{\alpha}j}V_{{\beta}j}^*V_{{\beta}i}
   \exp\left(i\varphi_{ij}-\mathscr{A}_{ij}^2-\mathscr{B}_{ij}^2\right),
\end{eqnarray}
but with the essential difference that the factors $\mathscr{A}_{ij}$ and $\mathscr{B}_{ij}$
do depend (through the functions $\mathfrak{D}$ and $\mathfrak{n}$) on the neutrino energy and momenta of the external wave packets.
This dependence drastically affects the magnitude and shape of these factors if at least some of the wave packets
have relativistic momenta (that is always the case in the contemporary neutrino oscillation experiments).
As follows from our analysis, for sufficiently small and/or hierarchically different momentum spreads $\sigma_\varkappa$,
the functions $\mathscr{A}_{ij}$ and $\mathscr{B}_{ij}$ may vary in many orders of magnitude through their multidimensional domain.

The factors $\exp\left(-\mathscr{A}_{ij}^2\right)$ (with $i{\ne}j$) suppress the interference terms in \eref{TransitionProbability_limit}
at the distances exceeding the ``coherence length'' 
\[
L_{ij}^{\mathrm{coh}}=1/({\Delta}v_{ij}\mathfrak{D})\gg |L_{ij}|
\qquad
({\Delta}v_{ij}=|v_j-v_i|),
\]
when the neutrino wave packets $\psi_i^*$ and $\psi_j^*$ are strongly separated in space
(due to the difference in their group velocities) and no longer interfere. Clearly $L_{ij}^{\mathrm{coh}}\to\infty$
in the plane-wave limit.

The suppression factors $\exp\left(-\mathscr{B}_{ij}^2\right)$ ($i{\ne}j$) work in the opposite situation
when the external packets in $\mathcal{S}$ or $\mathcal{D}$ (or in both $\mathcal{S}$ and $\mathcal{D}$) are strongly
delocalized (in the plane-wave limit -- uniformly distributed over the whole space). 
The gross dimension of the neutrino production and absorption regions in $\mathcal{S}$ and $\mathcal{D}$
is of the order of  $1/\mathfrak{D}$. The interference terms vanish if this scale is large compared to the
``interference length'' 
\[
L_{ij}^{\mathrm{int}}=1/(4{\Delta}E_{ij})=2L_{ij}/(\pi\mathfrak{n}).
\]
In other words, the QFT approach predicts vanishing of neutrino oscillations in the plane-wave limit.
In this limit, the flavour transition probability does not depend on $\overline{L}$, $E_\nu$ and neutrino masses and becomes
$
\sum_{i}|V_{{\alpha}i}|^2|V_{{\beta}i}|^2.
$
Thereby, a nontrivial interference of the diagrams with the intermediate neutrinos of different masses
is only possible if $\mathfrak{D}\ne0$. Our analysis of the generic subprocesses $1\to2$, $1\to3$ and $2\to2$ shows
that $\mathfrak{D}\ne0$ if in both vertices of the macrodiagram describing the process \eref{Macroprocess_A}
there are \emph{at least two} interacting wave packets $\varkappa$ (no matter in or out) with $\sigma_\varkappa\ne0$.
The same requirement unavoidably leads to the vanishing of the non-diagonal terms, when the mean distance between 
$\mathcal{S}$ and $\mathcal{D}$ becomes large enough in comparison with the coherence lengths $L_{ij}^{\mathrm{coh}}$.
As a result, the range of applicability of the standard QM formula for the neutrino oscillations probability is limited
by rather restrictive conditions,
\begin{eqnarray}
\label{QMapplicability}
\left\langle\left(\frac{2\pi{\mathfrak{D}}L}{E_{\nu}L_{ij}}\right)^2\right\rangle \ll 1
\qquad\mathrm{and}\qquad
\left\langle\left(\frac{{\pi}\mathfrak{n}}{2{\mathfrak{D}}L_{ij}}\right)^2\right\rangle \ll 1,                                     
\end{eqnarray}
in which the angle brackets symbolize an averaging over the phase subspace of the process \eref{Macroprocess_A}
which provides the main contribution into the measured count rate. 
We recall that the conditions \eref{QMapplicability}, as well as the relation \eref{TransitionProbability_limit},
were obtained under a number of assumptions and simplifications, which are not necessarily adequate to fully represent
the real-life experimental conditions.
Our consideration suggests that in the analysis and interpretation of real data, one should take into account
not only the conditions \eref{QMapplicability}, but also the operating times of the source and detector, their geometry
and dimensions, explicit form of the distribution functions of in packets and other technical details.

\section{\label{sec:conclusions} Conclusions}

In this paper we have considered a covariant theory of wave packets in which the free field states are constructed as
superpositions of one-particle Fock states and are reduced to the Fock states in the plane-wave limit.
The mean momentum and effective spatial volume of such a wave packet are exact integrals of motion,
in spite of the natural dispersion of the packet with time.
The mean position of the packet evolves along the classical trajectory.
As a simplest model, satisfying all requirements of the formalism, we have studied relativistic Gaussian packets (RGP).
The conditions are found which allow one to neglect the spreading of RGP (CRGP approximation) and use the latter
for description of the asymptotically free particle states in the $S$-matrix formalism of QFT.

The formalism has been implemented for the calculation of Feynman diagrams with massive neutrino exchange between
two macroscopically separated vertices and with the wave packets as external legs. 
As a typical generic example, we have studied the amplitude of a lepton flavor violating process or rather
a class of the processes \eref{Macroprocess_A}
with production of the two charged leptons $\ell_\alpha$ and $\ell_\beta$ in the distant parts of the macrodiagram. 
The amplitude is proportional to the sum of the products of the matrix elements describing the subprocesses
of neutrino creation and absorption and the factor $\psi_j^*/L$, which can be treated as a spherical wave of neutrino
$\nu_j$ with a mass $m_j$. The factor $\psi_j^*$ turns to be of the same form as the generic CRGP wavefunction,
but its ``dispersion'' is a relativistic-invariant function of the neutrino energy $E_\nu$ and momenta
of external wave packets $\boldsymbol{p}_\varkappa$.
As a common multiplier, the amplitude contains a suppression factor, arising due to the incomplete spacetime overlap of
the external wave packets and, in addition, two non-singular factors, responsible for an approximate energy-momentum conservation
in the vertices of the macrodiagram.

Experimental detection of the flavor-violating process \eref{Macroprocess_A} would provide a firm indication of the
incompleteness of the standard model, and a detailed measurement of its probability would be an important source of
information on the neutrino mixing parameters and squared-mass splittings.
A Lorentz-invariant formula for the probability of the process \eref{Macroprocess_A} has been obtained.
A macroscopic averaging of this probability leads to an experimentally measurable quantity --
the differential number of events represented, after several simplifications, by a multidimensional integral
of the product of three factors, see \eref{AveragedProbability_3_b}.
The factors $d\Phi_\nu$ and $d\sigma_{{\nu}\mathcal{D}}$ are respectively related, but not identical, to the differential
flux density (energy spectrum) of massless neutrinos from the source and to the differential cross section of
the neutrino-detector interactions.
The function $\mathcal{P}_{\alpha\beta}$ is answerable for the neutrino flavor transitions.
But in general this quantity does not have the properties of a probability, because
it involves the decoherence factors dependent on the phase space variables $E_\nu$ and $\boldsymbol{p}_\varkappa$ and on the
masses and momentum spreads of the external wave packets, including the packets of the leptons $\ell_{\alpha}$ and $\ell_{\beta}$.

Within a simple model for the distribution functions of in packets in the source and detector,
the magnitude of the common suppression factor $S_0$ in $\mathcal{P}_{\alpha\beta}$ is determined by the operating times of the source
($\tau_s$) and detector ($\tau_d$), and by the spacetime width of the neutrino wave packet $\tau_\nu\sim1/\mathfrak{D}$,
where $\mathfrak{D}$ is a function of $E_\nu$ and $\boldsymbol{p}_\varkappa$.
The suppression is small if $\tau_s\gg\tau_d$ or $\tau_s\sim\tau_d\gg\tau_\nu$.
In the opposite case, the strong dependence of $S_0$ on the parameters $\tau_s$ and $\tau_d$ provides a potential possibility
of \emph{measuring} an average value of the function $\mathfrak{D}$ in a special-purpose accelerator experiment, which allows
us to variate these parameters (or at least one of them). Such a measurement, even if rather rough, would be useful
in design and data-handling of future precision experiments with neutrino factories, $\beta$-beams and super-beams.
The non-diagonal decoherence factors $S_{ij}$ ($i{\ne}j$) show in general a very involved behavior.
In an asymptotic regime, $\tau_s\gtrsim\tau_d\gg\tau_\nu$, these factors cease to depend on $\tau_s$ and $\tau_d$ but
continue to strongly suppress  the ``oscillation'' terms in $\mathcal{P}_{\alpha\beta}$.
The conditions have been found at which the decoherence effects are small and the standard QM formula
for the neutrino flavour transition probability is applicable.

\ack

This work was supported by the RFBR grant no~07-02-00215-a and by the Federal Target Program
``Scientific and Scientific-Pedagogical Personnel of the Innovative Russia'', contract no~02.740.11.5220.
DN is partially supported by a JINR Research Fellowship for Young Scientists.
We thank Dmitry Kazakov, Slava Lee and Oleg Teryaev for useful discussions.


\section*{References}

\begin{thebibliography}{99}

\bibitem{Pontecorvo:57}
  Pontecorvo B 1957
  \emph{``Mesonium and anti-mesonium''},
  {\it Zh.\ Eksp.\ Teor.\ Fiz.} {\bf 33} 549 [{\it Sov.\ Phys.\ JETP} {\bf 6} 429] 
  \item[] 
  Pontecorvo B 1957
  \emph{``Inverse beta processes and nonconservation of lepton charge''},
  {\it Zh.\ Eksp.\ Teor.\ Fiz.} {\bf 34} 247 [{\it Sov.\ Phys.\ JETP} {\bf 7} 172]


\bibitem{Kayser:81}
  Kayser B 1981
  \emph{``On the quantum mechanics of neutrino oscillation''},
  {\it Phys.\ Rev.} D {\bf 24} 110
  
\bibitem{Giunti:91}
  Giunti C, Kim C W and Lee U W 1991
  \emph{``When do neutrinos really oscillate? Quantum mechanics of neutrino oscillations''},
  {\it Phys.\ Rev.} D {\bf 44} 3635

\bibitem{Giunti:93}
  Giunti C, Kim C W, Lee J A and Lee U W 1993
  \emph{``On the treatment of neutrino oscillations without resort to weak eigenstates''},
  {\it Phys.\ Rev.} D {\bf 48} 4310
  (arXiv:hep-ph/9305276)

\bibitem{Rich:93}
  Rich J 1993
  \emph{``The Quantum mechanics of neutrino oscillations''},
  {\it Phys.\ Rev.} D {\bf 48} 4318

\bibitem{Giunti-Kim:07}
  Giunti C and Kim C W 2007
  \emph{Fundamentals of Neutrino Physics and Astrophysics}
  (Oxford University Press Inc., New York) 

\bibitem{Akhmedov:09}
  Akhmedov E Kh and Smirnov A Yu 2009
  \emph{``Paradoxes of neutrino oscillations''},
  {\it Yad.\ Fiz.} {\bf 72} 1417 [{\it Phys.\ Atom.\ Nucl.} {\bf 72} 1363]
  (arXiv:0905.1903 [hep-ph])
  
\bibitem{Akhmedov:10a}
  Akhmedov E Kh and Kopp J 2010
  \emph{``Neutrino oscillations: Quantum mechanics vs.\ quantum field theory''},
  {\it JHEP} {\bf 04} 1
  (arXiv:1001.4815 [hep-ph])
  

\bibitem{Grimus:96}
  Grimus W and Stockinger P 1996
  \emph{``Real oscillations of virtual neutrinos''},
  {\it Phys.\ Rev.} D {\bf 54} 3414
  (arXiv:hep-ph/9603430)

\bibitem{Grimus:99}
  Grimus W, Mohanty S and Stockinger P 1999
  \emph{``Field theoretical treatment of neutrino oscillations: the strength of the canonical oscillation formula''},
  arXiv:hep-ph/9909341
  \item[] 
  Grimus W, Mohanty S and Stockinger P 1999
  \emph{``Field-theoretical treatment of neutrino oscillations''},
  arXiv:hep-ph/9904340
  \item[] 
  Grimus W, Mohanty S and Stockinger P 2000
  \emph{``Neutrino oscillations and the effect of the finite lifetime of the neutrino source''},
  {\it Phys.\ Rev.} D {\bf 61} 033001
  (arXiv:hep-ph/9904285)
  \item[] 
  Stockinger P 2000
  \emph{``Introduction to a field-theoretical treatment of neutrino oscillations''},
  {\it Pramana} {\bf 54} 203
  
\bibitem{Cardall:99b}
  Cardall C Y 2000
  \emph{``Coherence of neutrino flavor mixing in quantum field theory''},
  {\it Phys.\ Rev.} D {\bf 61} 073006
  (arXiv:hep-ph/9909332)

\bibitem{Ioannisian:98}
  Ioannisian A and Pilaftsis A 1999
  \emph{``Neutrino oscillations in space within a solvable model''},
  {\it Phys.\ Rev.} D {\bf 59} 053003
  (arXiv:hep-ph/9809503)

\bibitem{Beuthe}
  Beuthe M 2003
  \emph{``Oscillations of neutrinos and mesons in quantum field theory''},
  {\it Phys.\ Rept.} {\bf 375} 105
  (arXiv:hep-ph/0109119)
 \item[] 
  Beuthe M 2001
  \emph{``Propagation and oscillations in field theory''},
  Ph.\,D.\ Thesis (Universit\'e catholique de Louvain, Sep.\ 4, 2000)
  UCL-IPT-00-12
  (arXiv:hep-ph/0010054)
  \item[] 
  Beuthe M 2002
  \emph{``Towards a unique formula for neutrino oscillations in vacuum''},
  {\it Phys.\ Rev.} D {\bf 66} 013003
  (arXiv:hep-ph/0202068)

\bibitem{Raghavan:05-06}
  Raghavan R S 2005
  \emph{``Recoilless resonant capture of antineutrinos''},
  arXiv:hep-ph/0511191
  \item[] 
  Raghavan R S 2006
  \emph{``Recoilless resonant capture of antineutrinos from tritium decay''},
  arXiv:hep-ph/0601079

\bibitem{Akhmedov:08-09}
  Akhmedov E Kh, Kopp J and Lindner M 2008
  \emph{``Oscillations of M\"{o}ssbauer neutrinos''},
  {\it JHEP} {\bf 05} 005
  (arXiv:0802.2513 [hep-ph])
  \item[] 
  Akhmedov E Kh, Kopp J and Lindner M 2009
  \emph{``On application of the time-energy uncertainty relation to M\"{o}ssbauer neutrino experiments''},
  {\it J.\ Phys.} G {\bf 36} 078001
  (arXiv:0803.1424 [hep-ph])

\bibitem{Bilenky:08-09}
  Bilenky S M, von Feilitzsch F and Potzel W 2008
  \emph{``Time-energy uncertainty relations for neutrino oscillation and M\"{o}ssbauer neutrino experiment''},
  {\it J.\ Phys.} G {\bf 35} 095003
  (arXiv:0803.0527 [hep-ph])
  \item[] 
  Bilenky S M, von Feilitzsch F and Potzel W 2009
  \emph{``Reply to the comment on 'On application of the time-energy uncertainty relation to M\"{o}ssbauer neutrino experiments' by E.Kh.~Akhmedov, J.~Kopp and M.~Lindner''},
  {\it J.\ Phys.} G {\bf 36} 078002
  (arXiv:0804.3409 [hep-ph])

\bibitem{Kopp:09}
  Kopp J 2009
  \emph{``M\"{o}ssbauer in quantum mechanics and quantum field theory''},
  {\it JHEP} {\bf 06} 049
  (arXiv:0904.4346 [hep-ph])

\bibitem{Potzel:09}
  Potzel W 2009
  \emph{``M\"{o}ssbauer antineutrinos: some basic considerations''},
  {\it Acta Phys.\ Polon.} B {\bf 40} 3033
  (arXiv:0912.2221 [hep-ph])

\bibitem{Peskin:95}
  Peskin M E and Schroeder D V 1995
  \emph{An Introduction to quantum field theory}
  (Addison-Wesley Publishing Company)

\end{thebibliography}
\end{document}